\documentclass[twocolumn,tighten,twocolappendix]{aastex63}
\usepackage{amsmath,amssymb,bm}
\usepackage{times}
\usepackage{comment}
\usepackage{array}
\usepackage{color}
\definecolor{ForrestGreen}{rgb}{0.133,0.545,0.133}  
\definecolor{MetallicViolet}{rgb}{0.357, 0.039, 0.567}
\definecolor{tangerine_dark}{rgb}{0.75, 0.52, 0.0}   
\definecolor{BlueViolet}{rgb}{0.541, 0.169, 0.886}
\definecolor{DarkBlue}{rgb}{0., 0., 0.33} 
\definecolor{RoyalBlue}{rgb}{0.25, 0.41, 0.88} 
        \newcommand{\R}[1]{{\color{red}{#1}}}  \newcommand{\M}[1]{{\color{magenta}{#1}}}  \newcommand{\T}[1]{{\color{tangerine_dark}{#1}}}  \newcommand{\V}[1]{{\color{MetallicViolet}{#1}}}

\newcommand{\Bep}{B_\mathrm{ep}}
\newcommand{\Bet}{B_\mathrm{et}}

\newcommand{\Bpot}{B_\mathrm{pot}}
\newcommand{\Df}{D_\mathrm{f}}
\newcommand{\ncr}{n_\mathrm{cr}}
\newcommand{\hcr}{h_\mathrm{cr}}

\submitjournal{ApJS}
\received{2026 March 18} \revised{2026 July 16} \accepted{2026 August 10} \published{}
\shortauthors{Jun Chen et al.}
\shorttitle{} 

\begin{document}

\title{Parametric Study of the Torus Instability Threshold}

\author[0000-0003-3060-0480]{Jun Chen}
\affiliation{Purple Mountain Observatory, Chinese Academy of Sciences, Nanjing 210023, China}

\correspondingauthor{Jun Chen}
\email{chenjun@pmo.ac.cn}

\author[0000-0002-5740-8803]{Bernhard Kliem}
\affiliation{Institute of Physics and Astronomy, University of Potsdam, Potsdam 14476, Germany}

\author[0000-0003-4618-4979]{Rui Liu}
\affiliation{
Department of Geophysics and Planetary Sciences, University of Science and Technology of China, Hefei 230026, China}

\begin{abstract}
The torus instability of an arched current channel has been suggested to initiate and drive major solar and stellar eruptions. Its threshold, given by the critical decay index of the equilibrium external poloidal field (the so-called strapping field) at the position of the current channel, is insufficiently known. Here, we carry out a parametric numerical study of the threshold, employing the force-free Titov-D\'emoulin (TD) equilibrium of a line-tied partial toroidal current channel and flux rope. This addresses the scatter of the threshold about its canonical value, $n_\mathrm{cr}=3/2$. Values scattering in the range $n_\mathrm{cr}\approx$\,1--2 are typically found in numerical and observational studies of flux rope eruptions on the Sun. For zero external toroidal (guide, or shear) field and approximately semicircular geometry (corresponding to minimal photospheric line-tying), we find the threshold to lie in the theoretically expected range of $\approx$\,1--1.5. An external toroidal field introduces a strong stabilizing effect on the instability, raising the threshold up to $\sim$\,2.5, which can explain observational and numerical results above the canonical value. Line-tying is found to act as a stabilizer as well. We also consider the approximate threshold based on the potential field and find a very good agreement with the exact numerical value, provided the horizontal component perpendicular to the flux rope axis is used to approximate the external poloidal field.
\end{abstract}


\section{Introduction}\label{s:intro}

The torus instability of a current channel, or, equivalently, a magnetic flux rope, is considered to be a key process in solar eruptions, which are observed as prominence/filament eruptions, coronal mass ejections (CMEs), flares, and jets. This ideal magnetohydrodynamic (MHD) instability is regarded as a prime candidate for the initiation of filament/prominence eruptions and CMEs and for the main acceleration of the plasma frozen in the erupting magnetic flux. It causes the expansion in major radius, $R$, of an arched or fully toroidal current channel, carrying a net current and held in force-free equilibrium by an external poloidal (``strapping'') magnetic field. The equilibrium is unstable if the external poloidal field falls off with $R$ sufficiently rapidly. Then, the inward Lorentz force of the current channel in the external poloidal field decreases faster with increasing $R$ than the channel's outward Lorentz self-force, so a perturbation of the major radius grows. This represents a lateral kink in the current channel. In the course of an expansion, the minor radius, $a$, grows as well.

The spatial profile of the external poloidal field can be quantified by its dimensionless decay index, defined as 
\begin{equation}                                                         
n_R(R)=-\frac{\mathrm{d}\ln \Bep(R)}{\mathrm{d}\ln R}
\label{e:n_R}
\end{equation}
in toroidal geometry. The threshold of the torus instability has been found to be 
\begin{equation}                                                         
n_{R,\mathrm{cr}}=3/2  
\label{e:n_cr3/2}
\end{equation}
in the simplest case of a toroidal current channel of large aspect ratio ($R/a\gg1$) in a vacuum with zero external toroidal field, $\Bet=0$ \citep{Osovets1959, Bateman1978}. However, line-tied flux ropes in the source regions of solar eruptions represent partial tori with a broad range of length-to-height ratios, with varying and often rather moderate aspect ratios, that are embedded in magnetized plasma, and that are typically in the presence of an external toroidal (guide or shear) field component, $\Bet\ne0$. The modeling of solar eruptions as well as numerical treatments of the instability clearly show that the threshold depends on the parameters that quantify these more general conditions and tends to fall in the range $\sim\!1\mbox{--}2$ in most cases, as shown in Section~\ref{s:review}. Quite often, however, the threshold has been found to lie near the value of 3/2, which is thus considered to be its canonical value. 

In order to infer reliably whether the instability causes the onset and driving of solar eruptions, and in order to utilize it in physics-based forecasts of such events, 
it is necessary to fully understand and quantify the parametric dependence of its threshold.
So far, much of the work addressing the parametric dependence, summarized in Section~\ref{s:review},
has focused on the 
geometry of the current channel. The results for $\Bet=0$ tend to support the 
canonical values $n_\mathrm{cr}=1$ (of the function $n(h)$) for the two-dimensional (2D) case of a cylindrical current channel at a height $h$ above a plane \citep{vanTend&Kuperus1978} and of $n_{R,\mathrm{cr}}=3/2$ in the toroidal (3D) case. However, threshold values for realistic, partial tori in the corona, line-tied in the photosphere, are controversial, with some investigations suggesting values
$n_\mathrm{cr}\sim1\mbox{--}1.5$ and others suggesting $\ncr<1$, even $\ncr<0$ in part of the relevant parameter range. Moreover,
the observational indication of 
threshold values somewhat higher than
3/2 for many active-region eruptions 
and in some numerical simulations is not explained by any theoretical treatment so far.
Further effects can be of similar or even stronger importance than the geometry, primarily the effect of an external toroidal field, $\Bet\ne0$. The displacement of the current channel must do additional work in deforming this flux, which on the Sun is frozen in ambient plasma; this can strongly raise the threshold \citep{Kliem&al2014b}. Second, a flux rope above a plane can possess a pure O-type 
or a combined O- and X-type 
magnetic structure, whose respective thresholds were estimated to differ by $\sim$\,0.5 \citep{Kliem&Torok2006}. 
Finally, the effect of line-tying and the deviation from the assumption of large aspect ratio increase with increasing minor flux rope radius. 

In this paper, we focus primarily on the stabilizing effect of an external toroidal (guide) field in order to address threshold values $>\!3/2$. We use the analytical model of a bipolar solar active region by \citet{Titov&Demoulin1999} as the initial condition in zero-beta ideal MHD simulations, because it facilitates performing a systematic parametric study with full freedom in choosing the value of $\Bet$. Despite its considerable flexibility, the TD equilibrium also introduces limiting constraints that prevent us from realistically studying the effects of 2D vs. 3D geometry and limit the study of how line-tying and magnetic topology 
influence the threshold. These constraints are considered in detail below (Sections~\ref{s:equilibrium}, \ref{s:parameters} and \ref{s:results}). Some of them, especially the deviation of the subphotospheric return current from the image current, are due to the strictly toroidal geometry of the current channel. Others result from the requirement of stability against the helical kink mode; these are of general nature. The parametric study also allows us to evaluate the often-used approximation of the decay index computed using the potential field as an approximation of the external poloidal field. 

Previous work on the parametric dependence of the instability threshold is reviewed in Section~\ref{s:review}.
We introduce the numerical model in Section~\ref{s:numerical}, the initial equilibrium in Section~\ref{s:equilibrium}, and the parameter range to be considered in Section~\ref{s:parameters}. The methods for determining the threshold for each set of parameters are illustrated in Section~\ref{s:representative}, followed by the results of the parametric study in Section~\ref{s:results} and the conclusions and a discussion in Section~\ref{s:concl}. In the appendix we consider the assumption that a torus-unstable flux rope expands approximately self-similarly at the onset of the instability, which underlies some of the expressions used in this paper.

\section{Review of Parametric Dependence}\label{s:review}

The torus instability was first described as an axial mode of the tokamak fusion device, i.e., an axially symmetric expansion of a toroidal current channel in a \emph{vacuum} \citep{Osovets1959}; for an alternative derivation, see \citet{Bateman1978}. These authors found the threshold of the instability in terms of the 
decay index of the 
external poloidal field $\Bep(R)$, 
$n_{R,\mathrm{cr}}=3/2$ (Equation~\ref{e:n_cr3/2}). 
This mode is easily stabilized in the tokamak device by including sufficiently large coils that produce a poloidal field of much lower decay index at the position of the current channel. Additional stabilization is provided by the currents induced in the walls of the device by the displacement of the current channel. 

\citet{Titov&Demoulin1999} and \citet{Kliem&Torok2006} considered the instability in the context of solar eruptions. The latter authors provided a description of the instability for a current channel in \emph{ambient magnetized plasma}, which is similar to the vacuum case if reconnection of the external poloidal field is allowed in the hole of the torus, corresponding to the area between the rising flux rope and the photospheric boundary on the Sun. The reconnection allows the flux rope to effectively slide through the external field, creating the similarity to the vacuum case. 

\citet{Kliem&Torok2006}
demonstrated that the characteristic acceleration profile is consistent with the properties of both fast and slow CMEs; hence, the instability serves as a unifying mechanism for these observationally very different categories of eruption \citep{MacQueen&Fisher1983}. An analytical description was 
possible by assuming an axially symmetric toroidal current channel of large aspect ratio, $R/a\gg1$, so the line-tying of arched flux ropes in the solar photosphere, expected to exert a stabilizing effect, was not included. Using the simplifying additional assumption of approximately self-similar expansion, $R/a\approx\mathrm{const}$, at the onset of instability, \citet{Kliem&Torok2006} also derived a correction term to the critical decay index, which depends inversely on the logarithm of the aspect ratio, 
\begin{equation}                                                         
n_{R,\mathrm{cr}}=\frac{3}{2}-\frac{1}{4[\ln(8R/a)-2+l_\mathrm{i}/2]}.
\label{e:n_cr3/2-corr}
\end{equation}
Here, $l_\mathrm{i}$ is the internal inductance per unit length of the torus (e.g., $l_\mathrm{i}=1/2$ for uniform current density in the cross section of the current channel, $l_\mathrm{i}=1$ for the axis-peaked force-free distribution by \citealt{Lundquist1950}, and $l_\mathrm{i}=0$ if the current flows only in the surface of the channel). The correction term lowers the threshold by a small amount of order 0.1--0.25 for aspect ratios $R/a\sim10\mbox{--}2$, which include the values typically indicated by filament channels and prominence cavities on the Sun. 

The expansion
instability of a toroidal 
flux rope has an analog in the translationally invariant 2D 
case of a cylindrical flux rope running above a boundary plane (i.e., without line-tying). The vertical instability of such a rope has already been suggested to explain solar eruptions by \citet{vanTend&Kuperus1978} and \citet{Molodenskii&Filippov1987}. In the 2D case, the critical value of the function $n(h)$ is $n_\mathrm{cr}=1$. \citet{Demoulin&Aulanier2010} developed a description of the instability that unifies the cylindrical (2D) and toroidal (3D) 
cases, also without including line-tying. For the straight rope, they also derived a small correction term, which is similar in structure and magnitude to the one in Equation~(\ref{e:n_cr3/2-corr}) but positive, raising the threshold slightly. They suggested that typically $1.1<n_\mathrm{cr}<1.3$. From these results, one expects the threshold to depend on flux rope geometry, increasing from $n_\mathrm{cr}\gtrsim1$ for very flat (nearly 2D) ropes to $n_\mathrm{cr}\lesssim3/2$ for nearly semicircular ones. \citet{Demoulin&Aulanier2010} and \citet{Kliem&al2014a} also demonstrated that the torus instability is equivalent to the description of the 2D and 3D cases as a fold catastrophe, or loss of equilibrium, in catastrophe theory \citep{Priest&Forbes1990, Forbes&Isenberg1991, Forbes&Priest1995, LinJ&al1998}. 

The toroidal models of \citet{Kliem&Torok2006} and \citet{Demoulin&Aulanier2010} (as well as \citealt{LinJ&al2002}) can be thought of as a semitoroidal coronal flux rope that keeps this shape while expanding, i.e., its footprints shift across the solar surface. The other half of the torus provides the subphotospheric return current, which is an exact image of the coronal current. Although the relevant scale, given by the footpoint half-separation
of the flux rope axis, $\Df$, enters when an initial-value problem is considered in numerical simulations or when the properties of solar eruptions are studied as a function of source region size, the effect of line-tying is obviously not included. This has been addressed in the analytical work of \citet{Isenberg&Forbes2007}, \citet{Olmedo&Zhang2010}, and \citet{Filippov2021a} and in the semianalytical treatment by \citet{Alt&al2021}. The latter three groups studied how the threshold changes when a changing fraction of the torus extends above the boundary surface and the photospheric footpoints of the toroidal axis are fixed. They computed the critical decay index as a function of apex height $h$ above the photosphere,
\begin{equation}                                                         
n(h)=-\frac{\mathrm{d}\ln \Bep(h)}{\mathrm{d}\ln h},
\label{e:n_h}
\end{equation}
considered to be more relevant than $n_R(R)$ for a current channel above a boundary plane. The two functions are related by 
\begin{equation}                                                         
n(h)=\frac{h}{R}\frac{\mathrm{d}R}{\mathrm{d}h}n_R(R).
\label{e:transform}
\end{equation}
The factor $(h/R)\mathrm{d}R/\mathrm{d}h$ introduces a strong height dependence for $h\lesssim R$, especially if the so-called shifted-circle model is adopted for the geometry of the current channel, as done by all three groups. This model assumes that the shape of the current channel is that of a partial circle, passing through the fixed footpoints and the apex point not only initially but also during the expansion. Then, the midpoint of the circle shifts with varying $h$ (i.e. $\mathrm{d}R/\mathrm{d}h\ne0$) and the sign of $(h/R)\mathrm{d}R/\mathrm{d}h=[(h/D_\mathrm{f})^2-1]/[(h/D_\mathrm{f})^2+1]$ changes at $h=R=D_\mathrm{f}$. If $n_R(R)>0$ in the height range around $h\sim D_\mathrm{f}$, then $n(h\!<\!D_\mathrm{f})<0$, suggesting the unexpected
result that flat line-tied current channels ($h<D_\mathrm{f}$) can be unstable in \emph{upward-increasing} strapping fields. This subtle point must be borne in mind when considering some of the results in the literature. Generally, attention must be paid to the difference between $n_R(R)$ and $n(h)$. 

\citet{Olmedo&Zhang2010} applied the shifted-circle model to a partially submerged full torus (i.e., the subphotospheric return current path is not an image of the coronal current path, except at $h=D_\mathrm{f}$). For self-similar expansion of a full torus, $n_{R, \mathrm{cr}}=\mathrm{const}$ from Equation~(\ref{e:n_cr3/2-corr}) and the factor $(h/R)\mathrm{d}R/\mathrm{d}h|_{h=\hcr}$ exactly determines $n(h)$ and $n_\mathrm{cr}$. For their two alternative assumptions considered for the expansion of the minor radius, the factor still dominates $n(h)$. Overall, for their three versions of the assumption, they obtained
$n_\mathrm{cr}<1$ if the coronal torus section is a fraction of a full torus less than $\approx\!0.6\mbox{--}0.8$ and even $n_\mathrm{cr}<0$ for fractions less than $\approx\!0.4\mbox{--}0.5$.

\citet{Alt&al2021} obtained a roughly similar behavior of $n_\mathrm{cr}$ for a full torus geometry. However, for the case of a subphotospheric image current, which is the appropriate way to model the force-free equilibrium of a coronal current channel \citep{Kuperus&Raadu1974}, they estimated $n_\mathrm{cr}\!\sim\!0.75\mbox{--}1.1$ in the 
range $h/D_\mathrm{f}\lesssim1$, which is most relevant for solar eruptions. 

\citet{Filippov2021a} developed an analytical description that includes both an approximation of the image current 
and the subphotospheric toroidal section that completes a full torus, i.e., two subphotospheric return currents. The upward force on the coronal current channel is dominated by the field of the image current when the coronal current channel is very flat ($h\ll D_\mathrm{f}$) and by the subphotospheric toroidal section when most of the torus runs above the photosphere ($h\gtrsim3D_\mathrm{f}$). 
Therefore, the superposition of the two force components yields the canonical 2D and 3D thresholds, $\ncr=1$ and $\ncr=3/2$, respectively, to a good approximation in these limiting cases. However, in the range $h\lesssim D_\mathrm{f}$, both forces are comparable to each other, and their superposition roughly doubles the upward force on the coronal current channel. Small threshold values, $\ncr<0.5$, result in the range $0.2<h/D_\mathrm{f}<1.5$, including 
negative ones around $h\approx0.8D_\mathrm{f}$.

The above results are significantly limited by their underlying assumptions \citep{Olmedo&Zhang2010, Filippov2021a} or involve estimates for the transformation from laboratory to solar conditions \citep{Alt&al2021}. Therefore, it is quite likely that future work will yield modified results, with those by \citet{Alt&al2021} probably being much closer to the true thresholds than the other two. It is remarkable, however, that all three investigations found line-tying to \emph{lower} the stability of an arched coronal flux rope in the solar-relevant height range. For $h\lesssim D_\mathrm{f}$, partly also at heights slightly above $D_\mathrm{f}$, threshold values $\ncr<1$ were found, clearly below the range of values obtained in previous research for the free torus and even the cylindrical current channel. This is opposite to the stabilizing effect of line-tying on other kink instabilities \citep{Hood&Priest1981}. The origin of the reduced stability at $h<D_\mathrm{f}$ likely lies in the fact that the major radius of a line-tied rising loop must \emph{decrease} in this height range. Both the outward hoop force component and total Lorentz self-force scale approximately as $I^2/R$, where $I(h)$ is the toroidal current \cite[e.g.,][]{Myers&al2016}. Therefore, a decreasing $R$ for a line-tied loop yields the possibility that the outward force decreases less rapidly during the expansion than in the case of a free torus. This may lower the threshold, although part of the free magnetic energy is consumed for the bending of the loop's upper part. However, the upward bending of the loop just above its footprints is a stabilizing effect not yet included in the above three treatments. This effect should also increase with decreasing $h$ in the range $h\lesssim D_\mathrm{f}$ for the following reasons. 
The footprint area increases with decreasing $h$ (and with increasing minor radius $a$); consequently, the length ratio of the middle current channel section to the end (footprint) sections decreases. In the middle section, all field lines twist around the entire channel and the Lorentz self-force acts fully, whereas toward the ends, an increasing fraction of field lines merely arch over the channel axis, and the Lorentz self-force here can be largely canceled by the line-tied boundary during an expansion.
Therefore, the question of whether line-tying acts as a stabilizing or destabilizing effect on the torus instability is not yet clarified. Numerical simulations and solar data do not support the low threshold values $\ncr<1$ obtained for $h\lesssim D_\mathrm{f}$ in \citet{Olmedo&Zhang2010}, \citet{Filippov2021a}, and \citet{Alt&al2021}.

Numerical investigations of the instability suggest that its threshold may take values in an even broader range. While \citet{Torok&Kliem2007}, \citet{Aulanier&al2010}, and \citet{JiangC&al2013} obtained $n_\mathrm{cr}\approx3/2$, supporting the canonical threshold value, careful follow-up studies of different models of photospheric driving prior to instability onset suggested a range $n_\mathrm{cr}\approx1.3\mbox{--}1.5$ \citep{Zuccarello&al2015, Zuccarello&al2016}, in good agreement with \citet{Kliem&Torok2006} and \citet{Demoulin&Aulanier2010}. However, \citet{Fan&Gibson2007} and \citet{Fan2010} found significantly higher values of $n_\mathrm{cr}=1.9$ and 1.74, respectively. These higher values may be due to the pure O-type flux rope topology in their simulations, which prevents the immediate onset of reconnection, while the lower values above were mostly obtained for the O-X-type topology that supports an immediate onset. These topologies are characterized by a bald-patch separatrix surface (BPS) at the surface of the rope \cite[see][]{Titov&al1993, Gibson&al2004} and by a hyperbolic flux tube (HFT) under the rope \citep{Titov&al2002}, respectively. Note that \citet{Zuccarello&al2015, Zuccarello&al2016} created the latter topology above a photospheric bald patch that survived from the photospheric driving phase. The higher values also support the very approximate treatment of the instability in the absence of reconnection in \citet{Kliem&Torok2006} which indicated a threshold of $n_{R,\mathrm{cr}}=2$. An even higher threshold value of $n_\mathrm{cr}=2.5$ was obtained by \citet{An&Magara2013} in a simulation of flux rope emergence. By comparison with the canonical value of 3/2, these authors concluded that another process must have caused the eruption in their experiment. However, their ambient flux had a strong toroidal component, which acts to stabilize \citep{Kliem&al2014b}, so their critical decay index value does not rule out the instability. Data-constrained MHD simulations of solar eruptions also tend to yield threshold values slightly above those in \citet{Demoulin&Aulanier2010}: $n_\mathrm{cr}\approx1.5\mbox{--}1.75$ \citep{Kliem&al2013}, $n_\mathrm{cr}>1.5$ \citep{JiangC&al2018}, and $n_\mathrm{cr}\approx1.5$ \citep{Inoue&al2018b}. These slightly higher values may result from the line-tying, from the less coherent structure of the modeled flux rope compared to the idealized models employed in the above investigations, or from the presence of a weak external toroidal field component. 

Observational investigations of eruption onset on the Sun also yield a broad range for the threshold value, which, in most cases, is computed from a potential-field model. \citet{ChengX&al2013, ChengX&al2020} found a range $n_\mathrm{cr}\approx1.3\mbox{--}1.9$ with an average $\left<n_\mathrm{cr}\right>=1.6$ for 8 eruptions from active regions and $n_\mathrm{cr}\approx0.9\mbox{--}1.6$ with an average $\left<n_\mathrm{cr}\right>=1.2$ for six eruptions of quiescent filaments. These values support the above findings on the role of the geometry, because, on average, the erupting flux is closer to a section of a circle for active regions and closer to a straight structure for quiescent filaments. Values above the canonical 3/2 are also supported. \citet{Vasantharaju&al2019} found a range $n_\mathrm{cr}\approx0.8\mbox{--}1.3$, averaging to 1.0, for a small sample consisting mainly of intermediate prominence eruptions, with a few active-region events. \citet{zou19} obtained a much larger range $n_\mathrm{cr}\approx0.4\mbox{--}2.5$, averaging to 1.5, for 21 erupting active-region filaments; this very large range appears to be influenced by a lower accuracy. Statistical studies of over 80 quiescent filament eruptions suggested $n_\mathrm{cr}\sim1$ \citep{Filippov&Den2001, Filippov&Zagnetko2008}. However, a (still small) sample of quiescent filaments that have risen to a nearly semicircular shape before erupting (five documented cases so far) yielded a range $n_\mathrm{cr}\approx1.5\mbox{--}1.8$ \citep{Myshyakov&Tsvetkov2020, Rees-Crockford&al2020}. The latter results are broadly consistent with those of \citet{ChengX&al2020} for active regions. A much larger statistical study of 904 eruptive filaments (with the majority ($\approx\!65\%$) originating in the quiet Sun, only $\approx\!20\%$ originating in active regions, and the rest being intermediate filaments) yielded a range $n_\mathrm{cr}\approx0.7\mbox{--}2$ with an average $\left<n_\mathrm{cr}\right>=1.1$ \citep{McCauley&al2015}. The use of projected heights introduced a trend toward lower $n_\mathrm{cr}$ in this investigation. \citet{Filippov2021b} reported low threshold values in the range $\ncr=0.2\mbox{--}1$ for nine eruptions of quiescent prominences observed from Earth and the STEREO satellites in quadrature, claiming support for the low values obtained in his theoretical work \citep{Filippov2021a}. However, the threshold values were computed at the upper edge of the prominences, which systematically underestimates the true threshold because the relevant structure (magnetic axis or geometrical center of the prominence cavity) lies typically substantially above the visible prominence material \cite[e.g.,][]{Bak-Steslicka2016}.

While the above investigations used estimates of eruption onset height from imaging data, the following three used nonlinear force-free field (NLFFF) extrapolation to infer the existence of a coronal flux rope and its height at the time of a magnetogram taken shortly before the observed eruption onset. There are pros and cons to either of these methods, but extrapolation appears to yield less reliable height information overall, as the various extrapolation schemes are known to scatter strongly in their output \cite[e.g.,][]{Schrijver&al2006}. \citet{DuanA&al2019} inferred the decay index at the onset point of major eruptions (i.e., mostly from active regions) using NLFFF extrapolation to estimate the height and potential-field extrapolation to approximate the strapping field. They obtained a range $n_\mathrm{cr}=0.5\mbox{--}2.0$, with one outlier having $\ncr=2.7$. For $\ncr>1.3$, all eruptions evolved into a CME, and in this group most eruptions had a decay index in the range $\ncr=1.5\mbox{--}1.7$, consistent with the results in \citet{ChengX&al2020}. \citet{JingJ&al2018} and \citet{GuptaM&al2024} also studied major active-region eruptions, using a different NLFFF extrapolation scheme. Both obtained rather low decay index values in the ranges $0.2\lesssim \ncr \lesssim1.6$ (38 events) and $\ncr<0.8$ (10 events), respectively, which either cast doubt on the occurrence of the torus instability in most of their events or on the suitability of the extrapolation scheme for estimates of flux rope height.

Laboratory studies of the onset of torus instability yielded a critical value $n_\mathrm{cr}\sim0.8$ \citep{Myers&al2015, Myers&al2016, Myers&al2017}. However, recent work has demonstrated that the combined effect of the strong driving of the current through the erupting loop by the discharging capacitor bank, the fixed path of the return current, and the eddy currents induced in the walls of the device yields a systematic downward shift of the critical decay index relative to solar conditions by an \emph{estimated} value of $\sim$\,0.1--0.3 \citep{Alt&al2021}. Therefore, the range of thresholds found in these experiments translates to solar values in the range $n_\mathrm{cr}\sim0.9\mbox{--}1.2$, slightly higher than the results of their semianalytical model assuming a subphotospheric image current 
and closer to the solar ones that are expected in the absence of an external toroidal field.

Overall, these investigations provide support for the conjecture that torus instability is the onset mechanism of many, if not most, solar eruptions, but they also show very clearly that the critical decay index value scatters significantly in the range $n_\mathrm{cr}\sim1\mbox{--}2$, possibly even somewhat beyond this range. Only some of the underlying parametric dependencies of the critical decay index have been addressed so far, primarily the 2D (cylindrical) vs.\ 3D (toroidal) geometry of the current channel and flux rope. These do not explain threshold values above 1.5. Therefore, we focus primarily on the stabilizing effects of a guide field ($\Bet\ne0$) and of line-tying in the following investigation.

\section{Numerical Model}\label{s:numerical}

The torus instability in its basic form is determined solely by the Lorentz force, so the relevant equations are those of ideal MHD without gravity and thermal pressure. They are written as dimensionless conservation laws 
\begin{align}
\partial_t \varrho&=
                 -\bm{\nabla\cdot}(\varrho\,\bm{u})\,,      \label{eq_rho}\\
\partial_{t}(\varrho\,\bm{u})&=
- 
\bm{\nabla \cdot} (\varrho \, \bm{u}\, \bm{u})
+\bm{\nabla \cdot \mathsf{T}} 
 +\bm{J\times B}\,,  
\label{eq_mot}\\
\partial_{t}\bm{B}&= 
-\bm{\nabla\cdot}(\bm{u\,  B-B\,  u})\,, \label{eq_ind}\\
\bm{J}&\equiv\bm{\nabla\times B}\,,    \nonumber           \label{eq_cur}\\
 \bm{\mathsf{T}}&\equiv
{R_e}^{-1}
\varrho\,
 (\bm{\nabla\,  u}+(\bm{{\nabla\,  u}})^T-\frac{2}{3}\bm{\mathsf{I}\ \nabla\cdot u})\,. \nonumber
\end{align}
Here, $\bm{\mathsf{T}}$ is the viscous stress tensor, included for numerical stability, $\bm{\mathsf{I}}$ is the second-order unit tensor, the superscript $^T$ denotes the transposition for a second-order tensor, and $R_e$ is the fluid Reynolds number. 
The variables are normalized by the initial field strength $B_0$, density $\varrho_0$, and corresponding Alfv{\'e}n velocity $V_\mathrm{A0} = B_0 / \sqrt{\mu_0\varrho_0}$ at the apex point of the flux rope axis, whose height, $h_\mathrm{a}$, and corresponding Alfv\'en time, $\tau_\mathrm{A} = h_\mathrm{a}/V_\mathrm{A0}$, serve to normalize the coordinates. The further normalization quantities are derived from these. 

A Cartesian box of size $[-5,5]\times[-5,5]\times[0,10]$ is resolved by a stretched grid of $141\times315\times200$ points. The resolution is nearly uniform at $\Delta=0.02$ in the volume of the initial flux rope ($|x|<0.65$, $|y|<2.66$, $z<3.02$) and degrades toward the upper and lateral boundaries. Closed boundaries ($\bm{u} = \bm{0}$ at all boundaries) and an invariant magnetogram ($\partial B_z / \partial t = 0$ at $z = 0$) are implemented, 
resulting in a line-tying effect. 
The initial magnetic field $\bm{B}(\bm{x},t = 0)$ is chosen to be the analytical, approximately force-free flux rope equilibrium by \citet{Titov&Demoulin1999}, as detailed in Section~\ref{s:equilibrium}. The initial density is chosen as $\varrho(\bm{x},t = 0)= |\bm{B}(\bm{x},t = 0)|^{3/2}$; this yields a gradual decrease in the Alfv\'en velocity from the flux rope toward the boundaries, with the height profile above the rope being close to an active-region height profile of $V_\mathrm{A}$ inferred from observations \citep{Vrsnak&al2002}. The plasma is at rest initially, $\bm{u}(\bm{x},t = 0)=\bm{0}$. 

Equations~(\ref{eq_rho})--(\ref{eq_ind}) are integrated by a modified Lax-Wendroff scheme \citep{Torok&Kliem2003}, whose diffusive Lax step is replaced by artificial smoothing \citep{Sato&Hayashi1979}. This replaces the density and velocity components (here jointly denoted as $\xi$) at each time step and grid point $i$ by a weighted average of the neighboring points $j$ as follows: 
  $$\xi_i \to (1-c)\xi_i +c\, \sum_{j=1}^6 \xi_j/6\,.$$  
This lowers the numerical diffusion by a factor $c\ll1$ compared to the Lax averaging, which is given by $c=1$. For most runs, the smoothing coefficient is chosen as $c=0.05$ in $\{z=0\}$, linearly decreasing up to $z=0.3$, and constant at $c=0.005$ in $\{z\ge0.3\}$. The numerical stability of the runs with $B_\mathrm{et} = 0$, $D_\mathrm{f}=1.6$, and $a=0.70$, 0.75, and 0.80 required a doubling of $c$. No such averaging is applied to the magnetic field ($c=0$).

\section{Titov-D\'emoulin Equilibrium}\label{s:equilibrium}

\begin{figure}[t]                                                          
 \centering
 \includegraphics[width=1\linewidth]{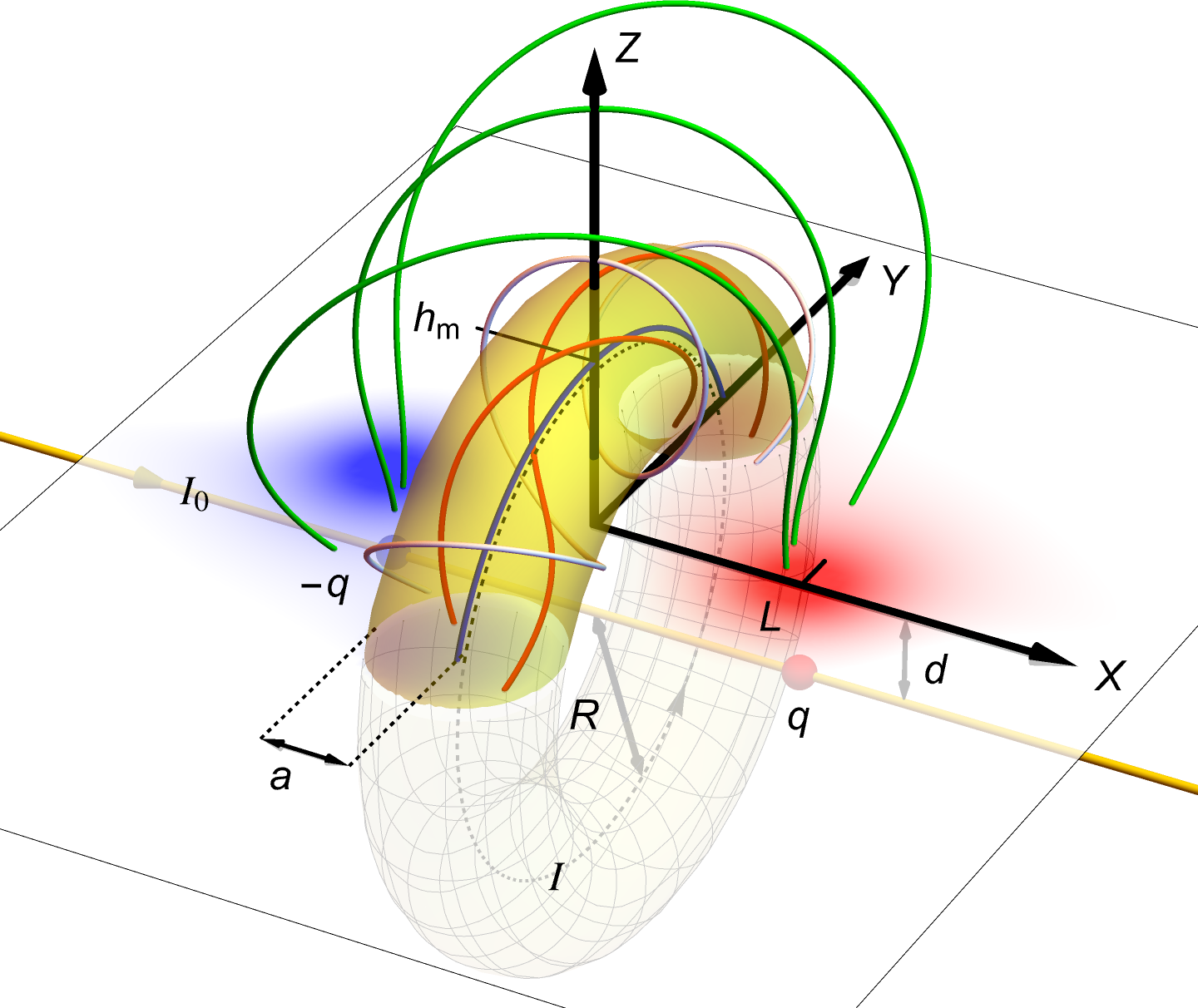}
 \caption{Initial (TD) equilibrium for $d/h_\mathrm{a}=0.57$, $a/h_\mathrm{a}=0.55$,
 	 $|\Bet/\Bep|=0.58$, where $h_\mathrm{a}=R-d$.
     The current channel is displayed through a transparent yellow volume rendering. 
 	 The blue field line is the magnetic axis of the flux tube; its apex height is $h_\mathrm{m}$. 
 	 The dotted line is the geometric axis of the torus at distance $R$
     from the torus center.
 	 Field lines of the flux tube are shown in red (inside the current channel) and pink (slightly outside the current channel).
 	 Green field lines display ambient flux. 
 	 The magnetogram, $B_z(x,y,0)$, is displayed in blue/red. } 
 \label{f:TD}
\end{figure}

The TD equilibrium \citep{Titov&Demoulin1999} is used as the initial condition of the simulations. It is constructed by superimposing the fields of a partially submerged toroidal current channel, a subphotospheric bipole, and a subphotospheric line current; the field in the current channel is finally matched to the ambient potential field at the surface. The current channel has major radius $R$ and minor radius $a$, and its center is submerged below the photospheric boundary, $\{z=0\}$, by a distance $d$, as illustrated in Figure~\ref{f:TD}. The magnetic charges of the bipole, $\pm q$, are positioned on the symmetry axis of the torus at $\bm{x}_{\pm}=(\pm L,0,-d)$. They provide the external poloidal (strapping) field, $\Bep$. The resulting magnetogram resembles that of a solar active region, so we will refer to $L$ as the ``sunspot distance.'' The equilibrium condition relates $q$ and the total ring (toroidal) current, $I$, to each other for a given geometry $(R, a, L)$. The line current runs along the symmetry axis of the torus and provides the external toroidal (guide/shear) field component, $\Bet$, which points exactly along the ring current, so its strength can be freely chosen. Because this choice for $\Bet$ decreases only slowly ($\propto R^{-1}$) with distance $R$ from its source, it prevents the instability from eventually developing into a CME \citep{Roussev&al2003, Torok&Kliem2005}, but this feature is not relevant at the onset of the instability studied here. 

The matching of the field in the current channel to the external potential field leads to a slight, aspect-ratio-dependent offset of the magnetic axis of the flux rope, whose radius, $R_\mathrm{m}$, is larger than the radius $R$ 
measured from
the center line of the current channel (``geometric axis'') by 7--24\% in our range of parameter values (increasing with decreasing aspect ratio $R/a$). The apex height of the geometric flux rope axis, $h_\mathrm{a}$ is chosen to be the length unit to avoid a dependence of length normalization on the thickness, $a$, of the current channel. 

The current channel carries purely forward (no return) current, as it is considered to be the most relevant for solar eruptions \citep{Torok&al2014, Dalmasse&al2015, LiuY&al2017, Kontogiannis&al2019, Vemareddy2019, LiuY&al2024}, so a strapping field is required for force-free equilibrium. The current density distribution is only weakly and smoothly varying in the volume of the current channel, with an enhancement in a thin layer at the surface from the matching to the external potential field. This is intermediate between the axis-peaked and ``hollow-core'' distributions of the extended (TDm) model \citep{Titov&al2014}. 

The magnetic flux rope is of the same geometry as the current channel, but is generally thicker (except if $d>R-a$, when there is no space for a current-free field between the current channel and the photosphere). The flux rope is bounded by a separatrix surface, or by a quasi-separatrix layer (QSL) if $\Bet\ne0$ and no bald patches exist \citep{Titov&Demoulin1999, Titov2007, Pariat&Demoulin2012}; we will collectively refer to these forms of flux rope boundary as a separatrix in the following. There are three cases that differ in the position and nature of the separatrix under the flux rope; these are also topologically different cases. When the separatrix touches the photosphere along the whole length of the flux rope, it does so in the polarity (PIL) and the horizontal field direction in this part of the PIL is then inverse (pointing from the negative to the positive polarity). Such sections of the PIL are referred to as bald patches and the corresponding separatrix, which is a true separatrix surface even for nonvanishing $\Bet$, is referred to as a BPS, see \citet{Titov&Demoulin1999} and, e.g., \citet{Gibson&Fan2006}. This configuration is of pure O-type. If the TD flux rope arches higher (for smaller depth $d$), the bald patch first splits in two, each with an attached BPS. These separatrices intersect each other under the rope in a field line called a separator or X-line, which introduces a combined O- and X-type structure \citep{Titov&Demoulin1999}. For an even higher arching rope, the two bald patches shrink and disappear, leaving a QSL (if $\Bet\ne0$, otherwise a separatrix surface) that intersects itself under the rope in an HFT of X-type structure \citep{Titov2007}. This configuration is of combined O- and X-type structure as well. 

Both the separator and HFT represent a seed for the formation of a vertical current sheet upon a displacement of the current channel, so fast reconnection can commence simultaneously with the displacement and the torus instability threshold should be near $n_\mathrm{cr}=3/2$ for $\Bet=0$ \citep{Kliem&Torok2006}. For a pure BPS flux rope, such a fast reconnection is inhibited by the line-tying in the bald patch until the displacement is large enough to cause internal current sheet formation by a lateral constriction of the upward-stretched flux rope, leading to its splitting \citep{Gibson&Fan2006}. Accordingly, existing studies \cite[especially][]{Kliem&Torok2006, Fan&Gibson2007, Fan2010} suggest a significantly higher threshold $n_\mathrm{cr}\lesssim2$ for $\Bet=0$, and possibly even higher for $\Bet\ne0$. However, \citet{Zuccarello&al2015} found far lower thresholds for the BPS case, very similar to the theoretical value for the HFT case, albeit with strong numerical diffusion permitting reconnection. The physics of the torus instability is not yet sufficiently understood in the BPS case \citep{Kliem&al2014a}, so the effect of topology may be different from the very approximate analytical result \citep{Kliem&Torok2006} and the above early numerical results. 

By its analytical nature, the TD model allows us to devise a broad parametric study. However, the model also comes with limitations. First, its expressions rely on the large-aspect-ratio approximation ($R/a\gg1$), so that the equilibrium is only approximate and attempts to relax to a true numerical equilibrium at the beginning of each MHD run. This changes the equilibrium from its nominal analytical values, the change increases with decreasing aspect ratio, and it can be quite significant. Second, and equally important, the geometry is strictly toroidal, implying that the subphotospheric return current is not an image of the coronal current. Specifically, the flatter the coronal current channel is chosen to be (by increasing $d$, which, in conjunction with the height normalization, implies a larger footpoint distance of the axis, $D_\mathrm{f}=(1+2d)^{1/2}$), the more distant is the return current, and correspondingly weaker is its field at the position of the coronal current channel. This implies that the 3D--2D transition of the instability threshold cannot realistically be studied. The return current is closest to the exact image current for nearly semicircular geometry ($d\ll1$). Additionally, the effect of line-tying can be studied only in a limited manner, because changing its strength (by changing $D_\mathrm{f}$) is always accompanied by a changing influence of the return current. 

While the initial TD equilibrium is of an exact toroidal shape as, e.g., in \citet{Olmedo&Zhang2010}, the numerical evolution of the current channel and flux rope does not follow the shifted-circle concept which assumes that the center of a circular current ring rises during the expansion. Rather, the center of the main (subphotospheric) part of the torus remains fixed and an image current arc moves downward (a consequence of the conservation of the normal field component in the bottom plane; see \citealt{Isenberg&Forbes2007}). Additionally, there is no need for the shifted-circle assumption in the numerical determination of the critical decay index. Therefore, in our analysis $\mathrm{d}R/\mathrm{d}h=1$ and $n(h)$ is always positive, differing from the results in \citet{Olmedo&Zhang2010} and \citet{Filippov2021a}.

\section{Parameter Range}\label{s:parameters}

Our goal is to determine the threshold $n_\mathrm{cr}$ of the torus instability for the following:
\begin{enumerate} 
\item varying external toroidal (guide) field strength (from $\Bet=0$ to $|\Bet|>|\Bep|$); 
\item varying aspect ratio $R/a$ extending significantly below values $\sim$\,10; 
\item both topological cases of pure O-type (BPS) and O-X-type (HFT) flux ropes. 
\end{enumerate} 
Although the TD model does not allow studying the geometrical (3D--2D) dependence of the instability threshold independently of other effects, we will nevertheless consider several values of the footpoint distance $D_\mathrm{f}$, because it turns out that the first two effects can best be studied for nearly semicircular shapes (small $d$ and $\Df$), but a transition between BPS and HFT flux ropes at small or zero guide field occurs only for large $\Df$. 

The external toroidal field strength is varied in the range $0\le|\Bet/\Bep|\le1.73$, where $\Bep$ is the equilibrium field strength of the analytical TD equilibrium at the magnetic axis of the current channel \cite[Equations~(4)--(6) in][]{Titov&Demoulin1999} and $\Bet$ is taken at the same position. Specifically, we choose $|\Bet/\Bep|=0,\,0.268,\,0.577,\,1.0,\,1.73$, which correspond, respectively, to the values $\psi=90$, 75, 60, 45, and 30~degrees for the photospheric projection of the angle between the external field, $(\Bep,\Bet)$, and the initial flux rope axis ($y$~axis) at the apex. The observations of flare loops and overlying/high-arching loops indicate that $\Bet$ is comparable to, and often even larger than, $\Bep$ in the field right above the flux rope, but decreases rapidly toward several initial flux rope (filament/prominence) heights. The external toroidal field in the TD equilibrium decreases only gently (linearly) with increasing distance from the line current, so its stabilizing effect is likely stronger than on the Sun for the same values of $|\Bet/\Bep|$ at the position of the flux rope. 

\begin{table}[t]                                                        
\centering
\caption{Range of parameter values considered in This study. Lengths are normalized by the apex height $h_\mathrm{a}=R-d$ of the geometrical flux rope axis. The normalized footpoint distances resulting from the selected range of $d$ are $D_\mathrm{f}=1.10\mbox{--}1.63$, and the toroidal aspect ratios lie in the range $R/a=1.4\mbox{--}3.3$. Dimensional values refer to $R=110$~Mm.}
\vspace{-2pt}  
\begin{tabular}{|c|c|c|}
\hline 
Parameter      & Dimensional Values    & Normalized Values \\
\hline 
$d$            & 10, 20, 30, 40, 50~Mm & .10, .222, .375, .571, .833 \\
$D_\mathrm{f}$ &                       & 1.10, 1.20, 1.32, 1.46, 1.63 \\
$a$            & 33--80~Mm             & .55, .60, .65, .70, .75, .80 \\
$|\Bet/\Bep|$  &                       & 0, .268, .577, 1.0,  1.73 \\ 
\hline
\end{tabular}
\label{t:parameters}
\end{table} 

Torus center depths are considered in the range of normalized values $0.1 \le d \le 0.833$, corresponding to normalized footpoint distances $\Df=1.1\mbox{--}1.63$. 
Here the smallest value is chosen as a compromise between a reasonable approximation of a subphotospheric image current and a not-too-unrealistic flux concentration in the ``sunspots.''
The largest value is close to the limit beyond which the Lorentz self-force, which depends on the distance of the return current (equivalently on $R$), is too weak to drive an unstable rise of the coronal current channel in the presence of line-tying. 

The minor radius $a$ influences the twist (and the topology) of the flux rope. Decreasing $a$ decreases the pitch of the field lines, because the force-free condition couples the radial and axial length scales of the rope; this increases the twist. We find that the twist remains subcritical with regard to the helical kink instability in the entire considered range of $d$ for normalized values of $a\ge0.55$. This limiting value yields a maximum twist of $\Phi=3.37\pi$ in our parametric study for $d=0.83$ and $\Bet=0$. No indications of the helical kink were observed in runs with these parameters. We choose $a=0.8$ as a reasonable maximum, which still leaves a little bit of space for potential field between the current channel and the photosphere. Flux ropes in the upper part of this range are representative of quiescent prominences, whose cavities indicate thick flux ropes even up to $a=1$ \citep{Forland&al2013}. 
The thinner ropes represent part of the range expected for eruptions from active regions, where extreme-ultraviolet hot channels suggest $a\sim0.2\mbox{--}0.5$ \cite[e.g.,][]{Reeves&Golub2011, ZhangJ&al2012, ChengX&al2014, YanXL&al2018}, although indications of thick flux ropes ($a\lesssim1$) exist as well \citep{Schrijver&al2008b}. 
\begin{figure}[t]                                                          
 \centering
 \includegraphics[width=.9\linewidth]{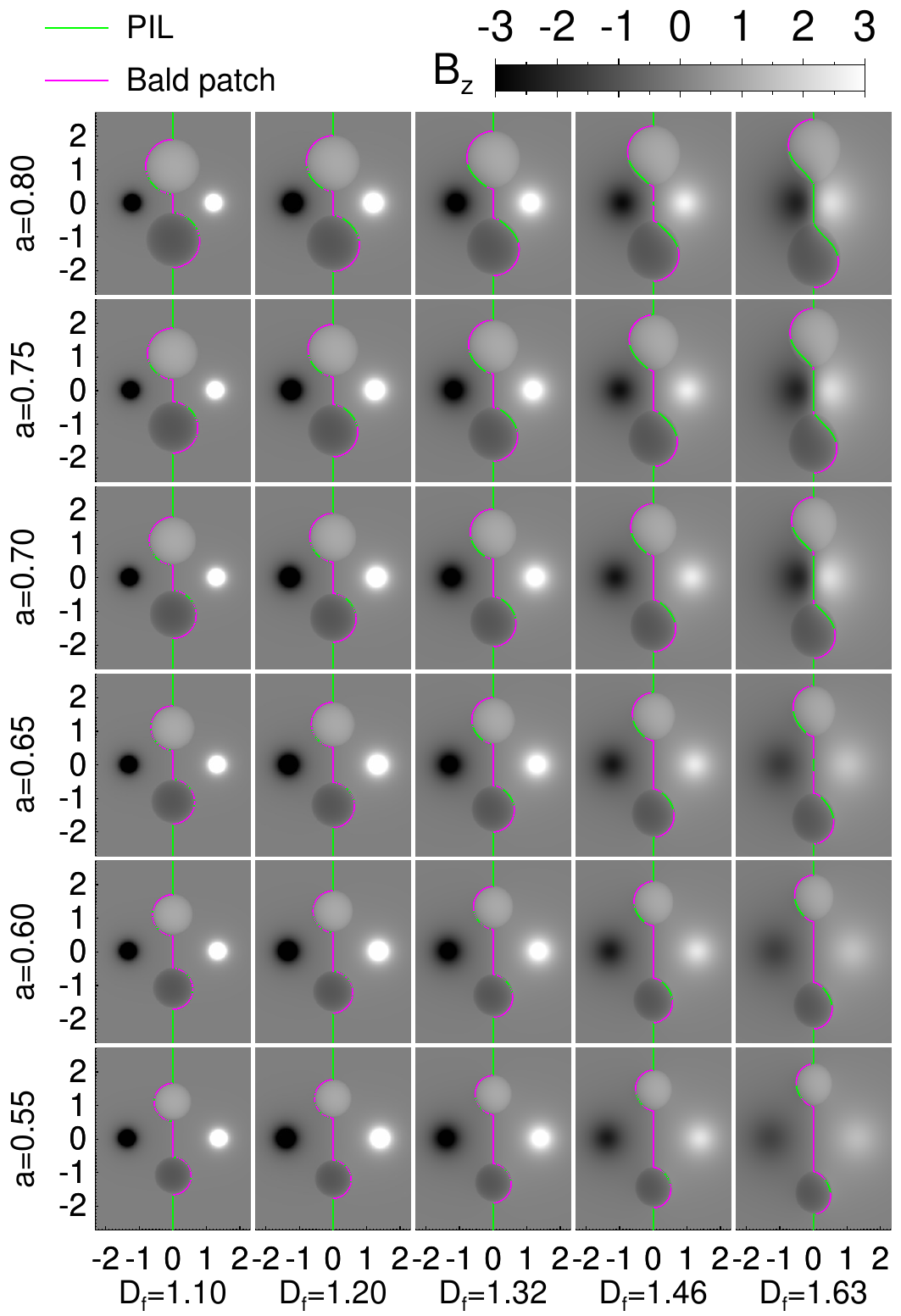}
 \caption{Bald-patch sections of the PIL of the marginally unstable analytical
  TD equilibria for $\Bet=0$.}
 \label{f:BP1}
\end{figure}

\begin{figure}[t]                                                          
 \centering
 \includegraphics[width=.9\linewidth]{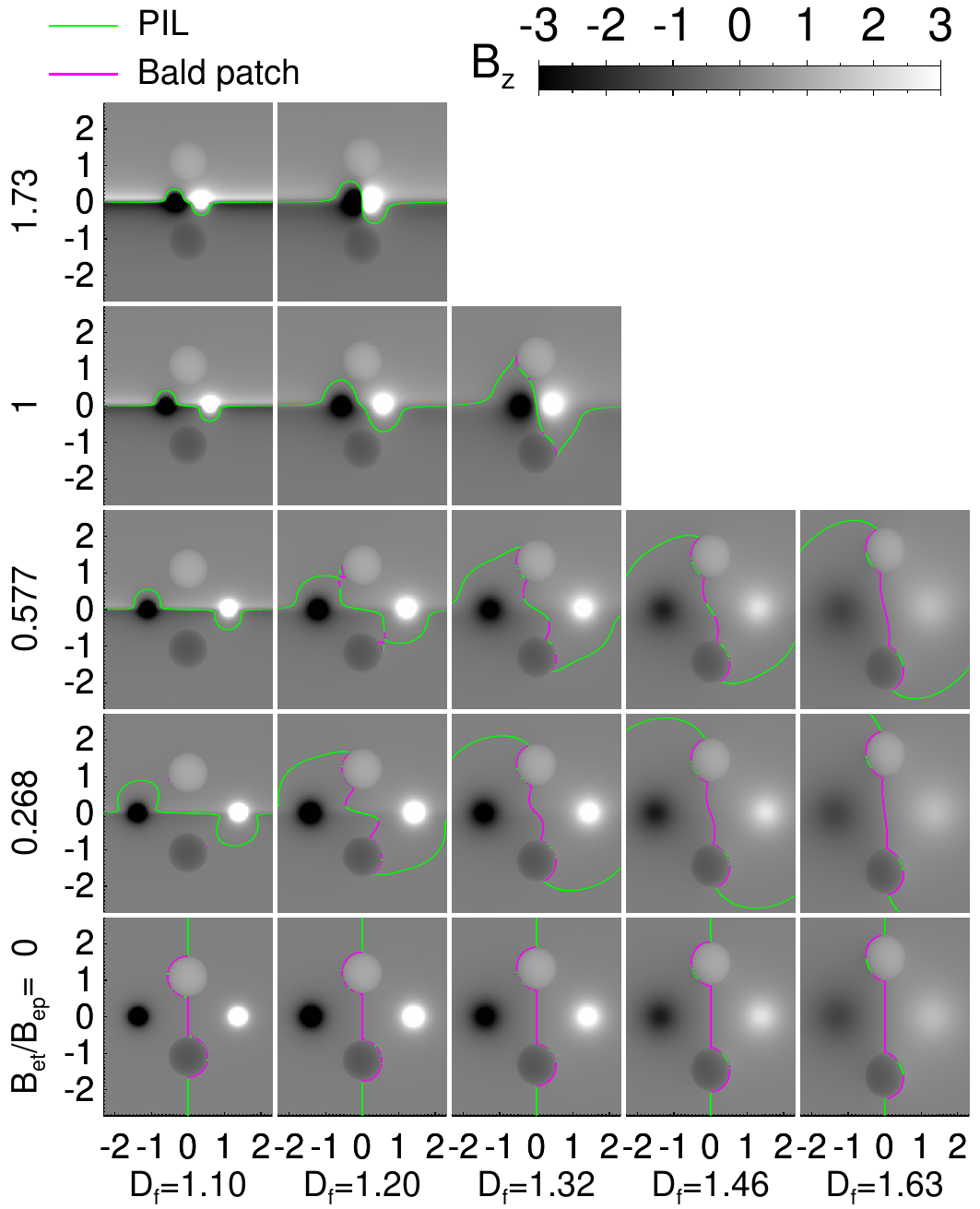}
 \caption{Transition of the marginally unstable analytical TD equilibria to split bald-patch and HFT structures with increasing $\Bet$, here for $a=0.55$.}
 \label{f:BP2}
\end{figure}

The range of our parametric study is compiled in Table~\ref{t:parameters}. As intended, the resulting toroidal aspect ratios extend to rather small values, $R/a=1.4\mbox{--}3.3$. For large values of $d$, the toroidal aspect ratio is of little relevance for the comparison with observations. Rather the ratio of the length to the minor radius of the coronal part of the current channel and flux rope is of interest. Using the arc length of the geometric axis of the current channel to characterize its length, this aspect ratio lies in the range $\approx$\,2.0--3.7. 

The topological change between pure O-type (BPS) and O-X-type (separator or HFT) flux ropes depends on the minor radius $a$ and the distance $L$ of the sources of $\Bep$: smaller $a$ and smaller $L$ lead to the transition from O-type to O-X-type. However, $L$ is \emph{not} a free parameter. 

We wish to determine the specific value $L_\mathrm{cr}$ that yields the marginally stable equilibrium for a given geometry, $\Df$ and $a$, of the TD flux rope and given ratio $|\Bet/\Bep|$.  For each set of these three parameters, we perform a series of simulations with varying sunspot distance $L$ to determine a lower and upper bound of the threshold value, $L_\mathrm{cr}$, for torus instability, i.e., the simulation study is \emph{four-dimensional} in parameter space. These bounds on $L_\mathrm{cr}$ translate straightforwardly to bounds on the critical height of the flux rope apex, $h_\mathrm{cr}$, and on the critical decay index, since $\Bep$ and its decay index on the $z$-axis are explicitly known (see Equation~(\ref{e:n_h_L}) below). The lower (upper) bounds on $L_\mathrm{cr}$ correspond to the upper (lower) bounds on $n_\mathrm{cr}$ and $h_\mathrm{cr}$.
We will refer to the configurations at the lower bound of $n_\mathrm{cr}$ as \emph{marginally stable} (this differs slightly from the use of the term for $n=n_\mathrm{cr}$ in the plasma physics literature) and to the ones at the upper bound to $n_\mathrm{cr}$ as \emph{marginally unstable}. The method for determining the marginally stable/unstable parameters will be detailed in Section \ref{s:representative}. 

For $\Bet=0$ it turns out that the marginally unstable TD equilibria possess a pure O-type structure for a large fraction of our chosen range of parameter values. Only the flattest and (surprisingly) the thickest flux ropes show the transition to an O-X-type structure (split bald patches with a separator for $(D_\mathrm{f},a)=(1.46,0.8)$ and $(1.63,0.65\mbox{--}0.75)$ and HFT for $(D_\mathrm{f},a)=(1.63,0.8)$). However, for our values of $\Bet>0$, most configurations are of pure HFT or split bald-patch type. Figure~\ref{f:BP1} shows the bald patches for $\Bet=0$ in the full range of our marginally unstable TD equilibria (also see the corresponding cross sections of toroidal current density at the flux rope apex 
in Figure ~\ref{f:jy}), 
and Figure~\ref{f:BP2} illustrates the disappearance of the bald patches with increasing $\Bet$ for $a=0.55$. One can also see that the modification of the magnetogram under the flux rope by the external toroidal field of the TD equilibrium leads to an unrealistic orientation of the PIL relative to the flux rope for the smallest chosen value of $d=0.1$ ($\Df=1.1$), which implies a weak $z$ component of the external poloidal field in this area. These configurations are discarded, so the dependence of the instability threshold on $|\Bet/\Bep|$ will be studied 
for $d=0.22$ ($\Df=1.2$), i.e., the remaining geometry closest to an exact subphotospheric image current. The effect of varying $a$ can be addressed for $d=0.1$. 

As the considered range of parameters is guided by solar observations of the aspect ratio and of stability against the helical kink mode in the majority of eruptions, the dominance of BPS equilibria is surprising, given the fact that solar observations indicate O-X-type configurations to be the typical case at eruption onset. This is inferred from the near-simultaneity of CME and flare onset \citep{ZhangJ&al2001, Neupert&al2001, ChengX&al2020}, which is facilitated by the presence of an X-type structure at eruption onset. The dominance of the BPS configuration for $\Bet=0$ in our parametric study is presumably related to the specific distribution of the current density in the current channel in a nontrivial way. The points of marginal stability and of transition between pure O and O-X structures tend to lie close to each other for a force-free flux rope equilibrium (see examples in \citealt{Forbes&Isenberg1991}, \citealt{Forbes&Priest1995}, \citealt{LinJ&al1998}, and \citealt{LinJ&al2002}), so small changes in the structure of the current channel may move the apex of the X-type structure above or below the photosphere. The strict toroidal geometry of the current channel also plays a role. Any X-type structure must have a downward curvature and thus tends to quickly reach down to the photosphere from its small apex height, corresponding to a split bald-patch configuration, while in 2D one would have a pure X-type configuration (note the short separation for two of the four split bald-patch configurations in Figure~\ref{f:BP1}).

\section{Representative Stable and Unstable Runs}
\label{s:representative}

\begin{figure*}[t]                                                          
	\centering
	\includegraphics[width=\linewidth]{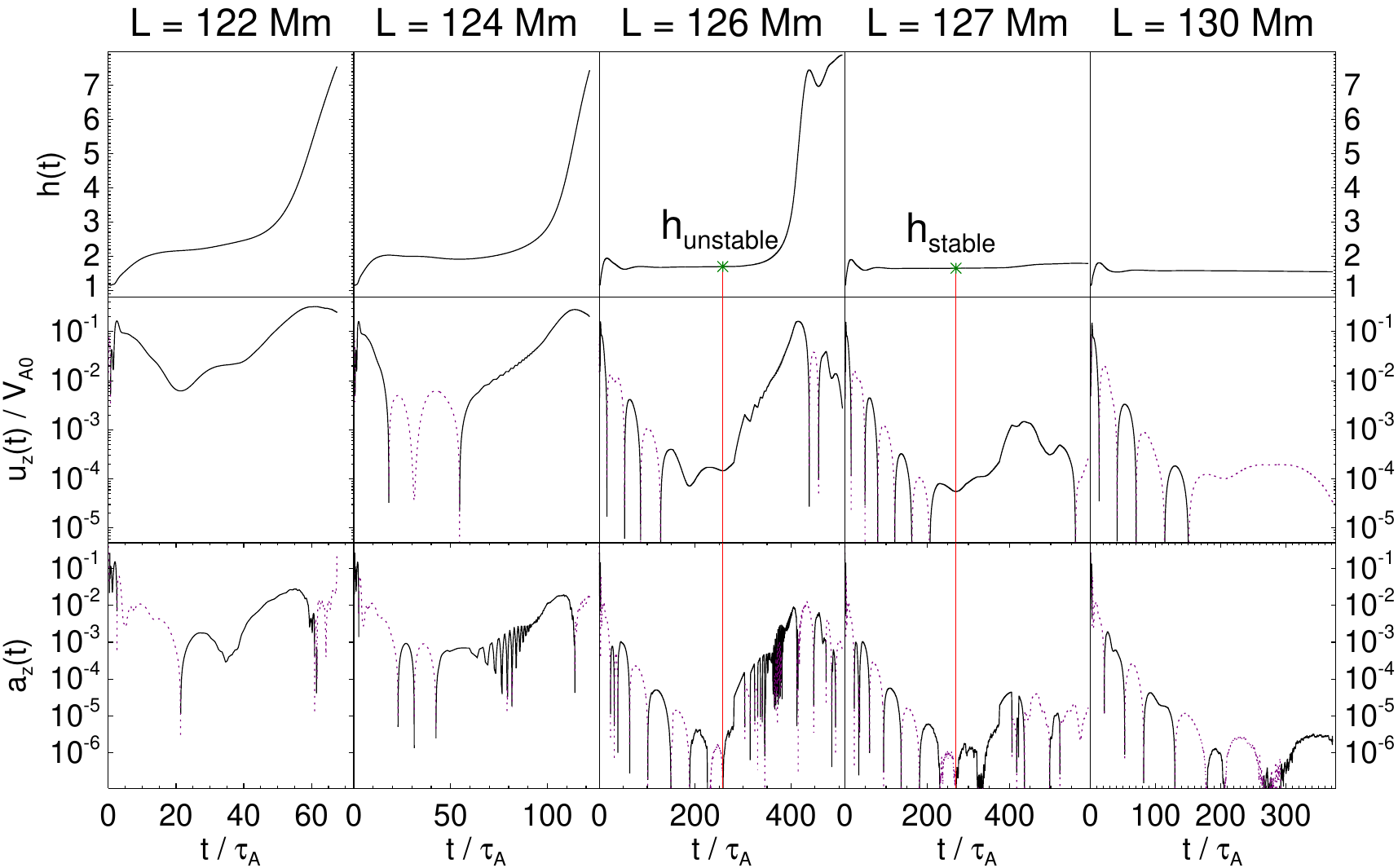}
	\caption{Temporal profiles of the motion of the fluid element at the apex point of the magnetic axis in the initial TD configurations for $a= 75$~Mm, $d= 10$~Mm, and $\Bet/\Bep = 0$. $L$ is varied from 122 to 130~Mm. The red line in the panel for $L = 126$~Mm (lower bound of $L_\mathrm{cr}$) marks the onset of instability; the red line in the panel for $L = 127$~Mm (upper bound of $L_\mathrm{cr}$) marks the end of the exponentially decreasing relaxation oscillations. 
	Dashed purple lines in the logarithmic plots of $u_z(t)$ and $a_z(t)$ show the absolute values of negative quantities.}
	\label{f:cfl}
\end{figure*}

For each value of $\Df$, $a$, and $|\Bet/\Bep|$, we estimate the critical sunspot distance $L_\mathrm{cr}$ and perform a series of runs with varying $L$ near $L_\mathrm{cr}$. The steps in $L$ between runs are reduced until the bounds on $L_\mathrm{cr}$ are small enough to finally yield small uncertainties in $\hcr$ and $\ncr$. To characterize each run, we monitor the height of the magnetic axis of the flux rope on the $z$-axis. The path of the fluid element at the apex point of the initial flux rope axis is integrated using the fluid velocity $u_z(z,t)$. Due to the line symmetry of the TD equilibrium about the $z$-axis, $B_{x,y}(-x,-y,z)=B_{x,y}(x,y,z)$ and $B_z(-x,-y,z)=-B_z(x,y,z)$, 
the fluid element remains at the $z$-axis. 

All runs begin with a relaxation from the initial approximate analytical equilibrium to a nearby numerical equilibrium, as shown for a specific parameter set in Figure \ref{f:cfl}. The stable configurations ($L>L_\mathrm{cr}$) show decaying relaxation oscillations about the numerical equilibrium (Figure~\ref{f:cfl} for $L\ge127$~Mm). The unstable configurations ($L<L_\mathrm{cr}$) begin with an equally quick  approach to the (unstable) numerical equilibrium, followed by an exponential, accelerated expansion (Figure~\ref{f:cfl} for $L\le126$~Mm). For values of $L$ very close to the lower bound of $L_\mathrm{cr}$, the instability has a small growth rate, so it develops very slowly, and relaxation oscillations about the numerical equilibrium can be seen before the exponential expansion begins to dominate (Figure~\ref{f:cfl} for $L=124\mbox{--}126$~Mm). 

For most parameter combinations $(\Df, a, |\Bet/\Bep|)$, the first unstable TD configuration in a sequence of runs with decreasing $L$ (the lower bound of $L_\mathrm{cr}$) is easily determined from the first appearance of an accelerated expansion into the upper part of the box. 
The onset time and onset height of the instability (the upper bound of $\hcr$) are then determined from the point of transition between the preceding relaxation oscillations about an equilibrium in the height range $h\sim1\mbox{--}2$ and the accelerated rise
(red line in Figure~\ref{f:cfl} for $L=126$~Mm). The transition is most clearly seen in the acceleration profile $a_z(t)$ (Figure~\ref{f:cfl}). We apply three criteria to determine the onset of instability: (1) The onset time is the transition point from oscillations with an overall decreasing amplitude to an overall increasing amplitude in $a_z(t)$, specifically between an oscillation half-period with a negative amplitude
and oscillations with a continuously increasing amplitude of the positive half-periods
(at such a transition point, the top part of the flux rope is closest to a force-free state). Ideally, the final oscillation half-period before the transition has the minimum amplitude in the series of relaxation oscillations. (2) The continuous increase in the amplitudes of the $a_z(t)$ oscillations after the transition point shows an approximately exponential behavior, and the roughly exponentially increasing trend also exists in $u_z(t)$. (3) $h(t)$ reaches a height $h>5$ at the end of the second stage (the subsequent evolution is already influenced by the approach of the top part of the flux rope to the upper boundary of the simulation box). The onset height of the instability (the upper bound of $\hcr$) is taken from $h(t)$ at the onset time thus inferred. The upper bound of $L_\mathrm{cr}$ is taken to be the next larger value of $L$ in the series of runs for the given $(\Df, a, |\Bet/\Bep|)$, and the corresponding lower bound of $h_\mathrm{cr}$ is given by the apex height of the numerical equilibrium in this run. 

Some runs do not adhere clearly to all three of these criteria: the final half-period of the relaxation can have positive acceleration $a_z(t)$, which means there is no zero-point between the relaxation and growth phases; the amplitude of the final relaxation oscillation does not always decrease clearly exponentially when very small velocities $u_z<10^{-3}$ have already been reached; some runs show high-frequency oscillations in $a_z(t)$ between the relaxation and growth phases. For these runs, the determination of the onset of instability is less definite. It is then guided by the requirement that an approximately exponential growth should persist until a height $h>5$ is reached. 

A stable run is expected to exhibit only decreasing relaxation oscillations. However, when the velocity has decreased to $|u_z|\sim10^{-4}$, the roundoff errors of the numerical scheme begin to dominate the acceleration and velocity, which are then no longer reliable. Their resulting irregular behavior can clearly be distinguished from the quite regular preceding relaxation oscillations (see Figure~\ref{f:cfl}, $L=127$~Mm at $t > 270\,\tau_\mathrm{A}$). The transition time is determined by employing the following criteria: (1) it is a zero-point between a negative and a positive oscillation half-period in $a_z(t)$; (2) the amplitude of this negative half-period must be smaller than that of all prior oscillations and smaller than one of following oscillations. The lower bound of $\hcr$ is then given by $h(t)$ at the transition time.

\section{Parametric Simulation Study}\label{s:results}

\subsection{Numerical Equilibrium}\label{ss:num_equil}

\begin{figure}[t]                                                          
	\centering
        \includegraphics[width=\linewidth]{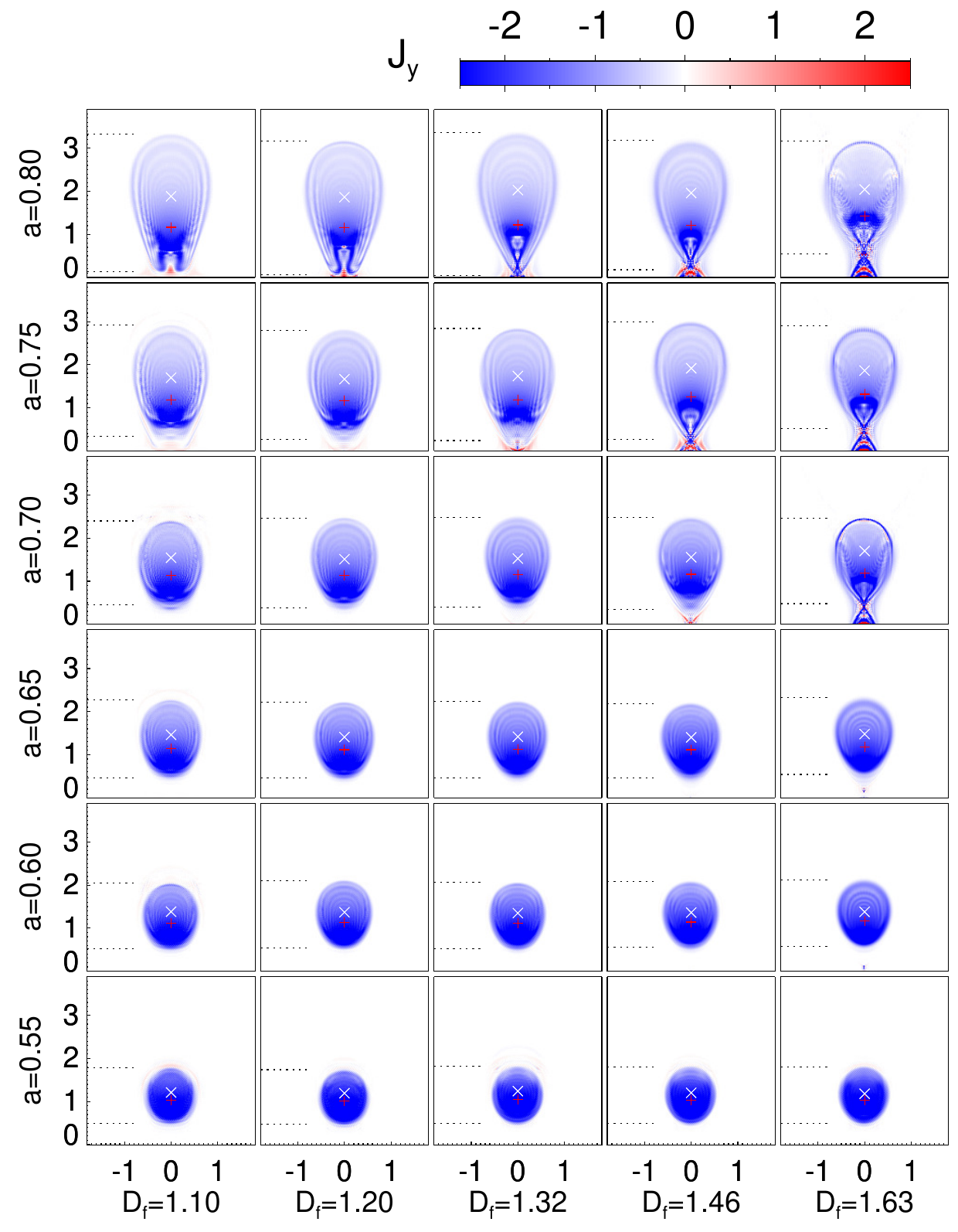}
	\caption{Toroidal current density $J_y(x,z)$ in the cross section, $\{y = 0\}$, of the marginally unstable current channel 
for $\Bet=0$, taken at the onset point of instability (similar to the red line at $L=126$~Mm in the third column of Figure~\ref{f:cfl}). A white cross marks the position of the flux rope's magnetic axis,  $z=h_\mathrm{m}$, and a red cross marks the effective apex height $\tilde{h}$ defined in Equation~(\ref{e:tilde_hcr}). 
Dotted lines mark the height range of the cross section.}
\label{f:jy}
\end{figure}

It turns out that the major flux rope radius expands in the initial relaxation from the analytical toward the numerical equilibrium in the entire range of considered parameter values. Since the analytical equilibrium uses the large aspect ratio approximation, $R/a\gg1$, the deviation of the numerical equilibrium from the analytical one increases with increasing $a$, reaching substantial values for our thick flux ropes. Figure~\ref{f:jy} shows the expanded numerical equilibrium state for all our marginally unstable configurations with $\Bet=0$ at the end of the initial relaxation phase by plotting the toroidal current density $J_y$ in the cross section of the flux rope at the apex. It is clearly seen that the expansion depends strongly on the thickness $a$ of the flux rope (or the aspect ratio $R/a$) but only weakly on the footpoint distance $D_\mathrm{f}$. While the expansion remains moderate for flux ropes of moderate thickness (our thinnest cases; $a=0.55$), it is very strong, approaching a doubling of the analytical equilibrium height for our thickest ropes ($a=0.8$). The expansion in minor radius is also nearly exclusively in the vertical direction, with mostly the upper half of the initial flux rope contributing to the expansion. The lower half, which contains most of the current, expands only moderately. This property of the expansion in minor radius is related to the relatively small values of $L_\mathrm{cr}$ at marginal stability in a large part of our parametric range. The smaller the $L$, the larger the change in the strapping field strength with height in the cross section of the current channel.

The fact that the numerical equilibrium is always reached by expansion shows that the analytical expression in the large aspect ratio approximation (Equation~5 in TD) overestimates the true Lorentz self-force of the current channel (averaged over the cross section). We have tried to correct for this mismatch by setting the total ring current somewhat below the analytical equilibrium value, but could reduce the initial expansion only by a small amount. A further reduction in the ring current resulted in a downward motion of the flux rope apex toward a significantly lower numerical equilibrium position. This is related to the specific height dependence of the current density (Equation~28 in TD), which yields a small net downward force in the lower part of the flux rope cross section and a small net upward force in the upper part. If the ring current is reduced too far below the analytical equilibrium value, then the net downward force dominates the relaxation. Consequently, we have worked with the analytical equilibrium value for the ring current. 

Some of the thicker and flatter ropes undergo a transition from a pure O to O-X structure in the course of relaxation.
By comparison with the bald patches in Figure~\ref{f:BP1}, one can see that at least two configurations ($a=0.7\mbox{--}0.75$ for $D_\mathrm{f}=1.5$) change from a pure O to O-X structure in the course of relaxation to the numerical equilibrium, bringing the number of HFT plus split bald-patch configurations to seven. Several additional cases very near a transition to an O-X structure exist among the neighboring configurations in these plots. From Figures~\ref{f:BP1} and \ref{f:jy} 
it is clear that the effect of BPS vs.\ HFT topology on the instability threshold can best be studied at the largest footpoint distances in our parametric range, $\Df=1.46\mbox{--}1.63$.

\subsection{Determination of Decay Index}\label{ss:ncr}

We primarily consider the decay index as a function of height, $n(h\!\equiv\!z)$ given by Equation~(\ref{e:n_h}), to provide a framework for comparison with estimates of the decay index from observational, numerical, and laboratory data. The corresponding value in toroidal geometry, $n_R(R)=(R/h)(\mathrm{d}h/\mathrm{d}R)n(h)$, will be considered in addition where helpful in understanding a particular result (Section~\ref{ss:results_line-tying}). The shifted-circle approximation will not be used in this transformation to avoid a sign change. Rather, we use the relation $h=R-d$, so that $\mathrm{d}h/\mathrm{d}R=1$.

Evaluating Equation~(\ref{e:n_h}) for the bipole of the TD equilibrium yields 
\begin{equation}                                                         
n(h)=\frac{3(d+h)h}{L^2+(d+h)^2}\,.  
\label{e:n_h_L}
\end{equation}
The bounds on the \emph{critical decay index} are taken to be the values of $n(h)$ at the numerically determined bounds on the critical height. In a first approach, we use the apex height of the flux rope's magnetic axis,
\begin{equation}                                                        
n_\mathrm{cr}:=n(h_\mathrm{cr})\,.      
\label{e:ncr_hcr}
\end{equation}

The current density redistribution toward the bottom part of the cross section for our values of $a\gtrsim0.7$, resulting from the initial expansion, is relevant for understanding the parametric dependence of the threshold, because most of the total current in these configurations runs at a height considerably below the magnetic flux rope axis. That is, the decay index $n(h_\mathrm{cr})$ 
refers to $\Bep$ significantly above the main part of the flux rope current, taking values that increasingly exceed the decay index in the height range of main current flow as $a$ increases. Therefore, in a second approach, we use an \emph{effective height}, $\tilde{h}$, appropriately weighted with the current density or Lorentz force across the cross section of the current channel at its apex. We regard this to be a more realistic measure of the critical height for onset of the instability. We have considered a weighting of the current density in the cross section of the current channel ($\tilde{h}$ being the center of gravity of $J_y(x,0,z)$) and a weighting of the Lorentz force in a wedge at the apex, with the tip of the wedge located at the torus center. Because the total Lorentz force is close to zero in the equilibrium configurations studied, one can use either the force of the strapping field or the Lorentz self-force as the weighting factor; both can be computed exactly for the TD equilibrium. It turns out that these weighting schemes yield very similar results for the effective height. We have finally applied the force-based weighting as follows:
\begin{equation}   
\tilde{h}:=\dfrac{\int_{V} z J_y(\bm{x})\Bep(\bm{x})\mathrm{d}\bm{x}} 
                 {\int_{V}   J_y(\bm{x})\Bep(\bm{x})\mathrm{d}\bm{x}}\,, 
\label{e:tilde_hcr}
\end{equation}
where $V$ is the volume of the wedge 
bounded in $z$ by two dashed lines in Figure~\ref{f:jy}. These effective heights are marked by the red crosses in Figure~\ref{f:jy}. The corresponding bounds on the \emph{effective critical decay index} are defined as 
\begin{equation}                                                        
\tilde{n}_\mathrm{cr}:=n(\tilde{h}_\mathrm{cr})\,.
\label{e:ncr_tilde_hcr}
\end{equation}

Different from the analytical model of the coronal field employed here, the external poloidal field is usually not known when the decay index is to be determined for observational or numerical data, because it is nontrivial to disentangle the poloidal field components from the current channel and from external sources. The horizontal component of the potential field, $\Bpot$, computed from the usually known magnetogram, is often 
used to approximate $\Bep$ at the position of the flux rope. This is based on the rationale that, for $\Bet=0$, the potential field typically crosses the PIL in projection in the perpendicular direction, as does $\Bep$ because the rope typically follows the PIL closely in projection. For $\Bet\ne0$, one can use the horizontal component perpendicular to the flux rope axis to approximate $\Bep$. The TD equilibrium allows us to assess the quality of this approximation by comparing $n_\mathrm{cr}$ (Equation~\ref{e:ncr_hcr}) with
\begin{equation}                                                        
\hat{n}_\mathrm{cr}:=-\left.\frac{\mathrm{d}\ln B_\mathrm{pot}(h)}{\mathrm{d}\ln h}\right|_{h=h_\mathrm{cr}}, 
\label{e:n_h_Bpot}
\end{equation}
or similarly, by comparing $\tilde{n}_\mathrm{cr}$ (Equation~\ref{e:ncr_tilde_hcr}) with this expression evaluated at $h=\tilde{h}_\mathrm{cr}$. Here we will compute both $\hat{n}_\mathrm{cr,hor}$ using $B_\mathrm{pot,hor}=(B_{\mathrm{pot},x}^2+B_{\mathrm{pot},y}^2)^{1/2}$ and $\hat{n}_\mathrm{cr,pol}$ using $B_\mathrm{pot,pol}=B_{\mathrm{pot},x}$. The TD rope axis initially aligns exactly with the $y$-axis and shows only little writhing (apex rotation about the $z$-axis) during the initial relaxation for our chosen small twist values, so the poloidal direction is nearly aligned with the $x$-axis. For $\Bet=0$ this is also perpendicular to the PIL under the rope (Figure~\ref{f:BP1}). For $\Bet\ne0$ the PIL under the rope apex follows the $y$-axis only approximately (Figure~\ref{f:BP2}), but the flux rope is still very well aligned with the $y$-axis after the numerical relaxation. Therefore, $B_{\mathrm{pot},x}$ approximates $B_\mathrm{pot,pol}$ very well in the entire range of parameters studied. 

The potential field can be computed using the Green function with the known normal field component in the bottom boundary, assuming zero field in this plane outside the box, as in extrapolations of observed magnetograms. However, since only $\mathrm{d}\Bpot/\mathrm{d}h$ is needed in our study of the decay index, this derivative is computed directly using the derivative of the Green function, which is more precise and efficient than the numerical differentiation of the computed $\Bpot$. For comparison, we also compute the potential field using the Fourier transform, as this method is often employed in the literature.

\subsection{Dependence on Minor Radius $a$ and Aspect Ratio $R/a$}\label{s:dependence_a}

First we consider the results for zero external toroidal field, $\Bet=0$. These can be best compared with the existing analytical results, many numerical results, and the observational results for quiescent filaments, which tend to be closer to $\Bet=0$ than flux erupting from active regions. Figure~\ref{f:n(D,a)} presents an overview of the critical heights and decay index values for our range of geometrical parameters $D_\mathrm{f}$ and $a$. The considerable expansion in major radius of the analytical TD current channel from $h_\mathrm{m}\approx1$ to a numerical equilibrium at $h_\mathrm{m}=h_\mathrm{cr}$ is seen to depend strongly on $a$ but only weakly on $D_\mathrm{f}$. For our thinnest flux ropes ($a=0.55$), the expansion from the analytical flux rope axis stays within 15\%, showing that the large aspect ratio approximation for the field of the toroidal current channel yields a reasonable force balance in the TD equilibrium down to the range of rather small values, $R/a\approx2\mbox{--}3.3$, realized for $a=0.55$. This is in basic agreement with the results in \citet{Zic&al2007}. For our thickest flux ropes ($a=0.8$, $R/a=1.4\mbox{--}2.3$), the initial height nearly doubles, and the effective height of the current channel, $\tilde{h}_\mathrm{cr}$, differs considerably from the magnetic axis height, $h_\mathrm{cr}$. 

\begin{figure}[t]                                                          
 \centering
 \includegraphics[width=\linewidth]{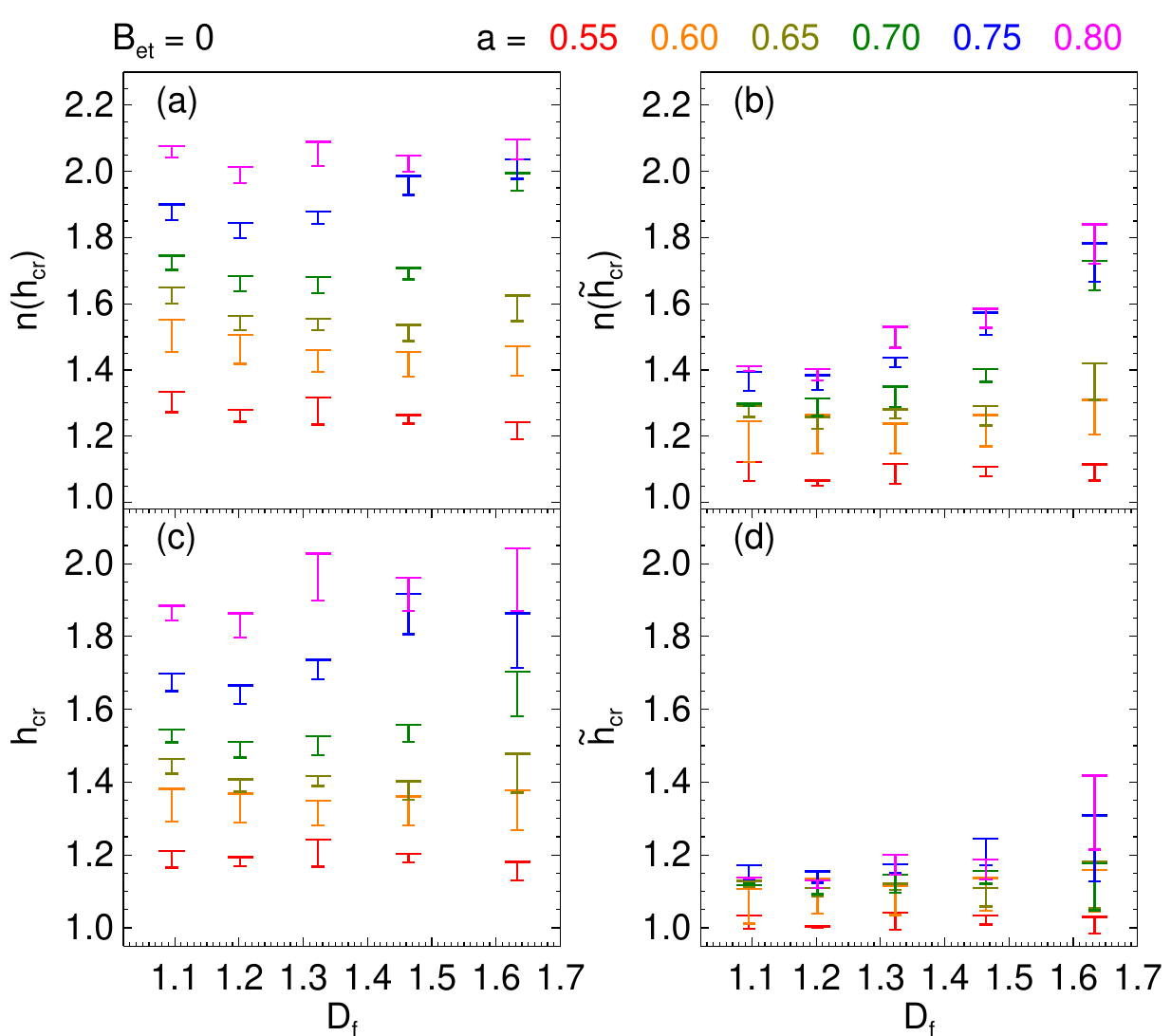} 
 \caption{Critical height and decay index as functions of geometry ($D_\mathrm{f}$ and $a$) for $\Bet=0$. The lower (upper) bounds on the critical values are shown as shorter (longer) horizontal bars. 
(a) Critical decay index at the magnetic axis, $n_\mathrm{cr}=n(h_\mathrm{cr})$. 
(b) Effective critical decay index at the Lorentz-force-weighted effective height,
    $\tilde{n}_\mathrm{cr}=n(\tilde{h}_\mathrm{cr})$. 
(c) Critical height of the magnetic axis, $h_\mathrm{cr}$. 
(d) Critical Lorentz-force-weighted effective height of the current channel, $\tilde{h}_\mathrm{cr}$ (Eq.~\ref{e:tilde_hcr}). 
} 
 \label{f:n(D,a)}
\end{figure}

\begin{figure*}[t]                                                          
 \centering
 \includegraphics[width=.8\linewidth]{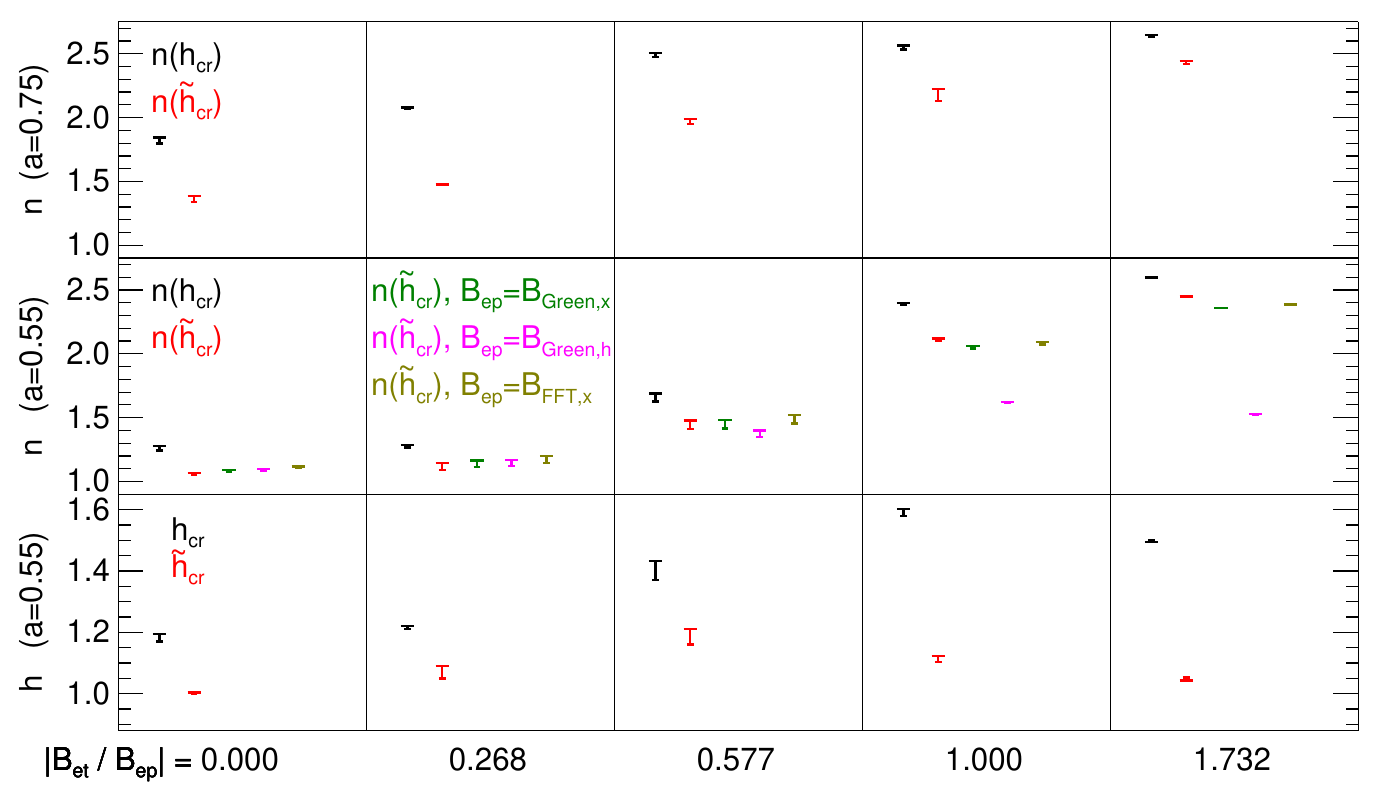}
 \caption{Critical height $h_\mathrm{cr}$, effective critical height $\tilde{h}_\mathrm{cr}$, and corresponding decay indices, $n_\mathrm{cr}$ and $\tilde{n}_\mathrm{cr}$, as functions $|\Bet/\Bep|$ for $\Df=1.2$ and $a = 0.55$ (bottom two rows), and critical decay indices for $a=0.75$ (top row). 
} 
 \label{f:n_bet}
\end{figure*}

We consider the effect of varying $a$ on the instability threshold for the smallest torus center depth ($d=0.1$; $\Df=1.1$) because the subphotospheric return current of the TD equilibrium is then closest to the image of the coronal current channel. The threshold values for $a=0.55$, $\tilde{n}_\mathrm{cr}\approx1.1$ and $\ncr\approx1.3$, lie in the range of values found in most theoretical and numerical investigations for $\Bet=0$, i.e., $\ncr(h)\sim1\mbox{--}1.5$, as summarized in Section~\ref{s:review} \cite[e.g.,][]{Kliem&Torok2006, Torok&Kliem2007, Demoulin&Aulanier2010, Aulanier&al2010, JiangC&al2013, Zuccarello&al2015, Zuccarello&al2016, Alt&al2021}. In particular, Equation~(\ref{e:n_cr3/2-corr}) yields $n_{R,\mathrm{cr}}=1.26$ for $a=0.55$, $\Df=1.1$, and $l_\mathrm{i}=0.5$, which transforms to $\ncr(h)=1.14$. 
These results support each other. 

The threshold increases when $a$ increases into the range where the large aspect ratio approximation is increasingly violated. The values at the magnetic axis of the rope are above the theoretically expected range of $\approx$\,1--1.5 for all $a\ge0.65$. We consider them to be increasingly unreliable. In contrast, the threshold values at the effective height $\tilde{h}$ fall into the theoretically expected range for all considered values of $a$: $\tilde{n}_\mathrm{cr}(h)\approx1.1\mbox{--}1.4$. This supports our definition of the effective height.

The increasing trend of $\ncr(h)$ and $\tilde{n}_\mathrm{cr}(h)$ with increasing $a$ is opposite to the trend in Equation~(\ref{e:n_cr3/2-corr}). However, 
the trend in Equation~(\ref{e:n_cr3/2-corr}) was found in the range of validity of the large aspect ratio approximation ($a\lesssim0.55$), 
while the trend found here (Figure~\ref{f:n(D,a)}) is obtained in the range of larger $a$. 

Figure~\ref{f:n(D,a)}(b) shows that the dependence of $\tilde{n}_\mathrm{cr}$ on $\Df$ for a fixed $a\le0.65$ is weak. The threshold values only increase significantly with increasing $\Df$ for $a\ge0.7$ and $\Df\ge1.5$. This increase is found in Section~\ref{ss:results_topology} to be due to a topology change. Otherwise, the weak dependence of $\tilde{n}_\mathrm{cr}$ on $\Df$ results from the superposition of three effects. As $\Df$ increases, so too does the distance to the subphotospheric return current in the TD equilibrium (hence, the driving force of the instability decreases), the footprint area of the TD rope (likely increasing the line-tying), and the aspect ratio $R/a$. The latter effect can be separated by considering larger normalized values of $a$ joint with larger values of $\Df$ in Figure~\ref{f:n(D,a)}, such that $R/a$ stays approximately equal,  
for example, from $(D_\mathrm{f},a)=(1.1,0.55)$ ($R/a=2$) to $(D_\mathrm{f},a)=(1.46,0.8)$ ($R/a=1.96$). This shows, as expected, that the threshold increases when the subphotospheric return current is located further away 
and the flux rope has a larger footprint area. The latter aspect will be considered in more detail in Section~\ref{ss:results_line-tying}.

\subsection{Stabilizing Effect of External Toroidal Field, $\Bet \ne 0$ }\label{ss:results_Bet}

Figure~\ref{f:n_bet} shows the critical heights and decay index values for $a=0.55$ and the critical decay indices for $a=0.75$ as a function of $|\Bet/\Bep|$, both for $\Df=1.2$ ($d=0.22$). This value of $d$ makes the subphotospheric return current as close as possible to the image of the coronal current channel while avoiding an unrealistic direction of the PIL under the flux rope (Figure~\ref{f:BP2}), and the smaller value of $a$ adheres best to the large aspect ratio approximation in our range of $a$ values. The initial expansion of the TD flux rope toward the numerical equilibrium, as measured by the magnetic axis apex height $\hcr$, increases strongly with increasing guide field strength $\Bet$, even for our smallest value of $a=0.55$, but the effective height $\tilde{h}$ at this minor radius depends only weakly on $\Bet$. The critical decay indices (both $\ncr$ and $\tilde{n}_\mathrm{cr}$) demonstrate the expected very strong stabilizing effect of the guide field in the whole considered range of minor radii $a$. 

This is likely the main reason for observational estimates of the critical decay index above 1.5 (in the range $\sim\!1.5\mbox{--}2.0$) for several eruptions from active regions \citep{ChengX&al2013, ChengX&al2020, DuanA&al2019, zou19} as well as from the quiet Sun \citep{Myshyakov&Tsvetkov2020, Rees-Crockford&al2020}. It also yields a plausible explanation for the high value of $\ncr=2.5$ in the simulation of flux emergence in \citet{An&Magara2013}. A strong guide field is conceivable especially in active regions when the unstable filament channel extends into the periphery of a sunspot, which is not uncommon. Our parametric study yields thresholds $\tilde{n}_\mathrm{cr}>2$ for flux ropes of moderate thickness $a=0.55$ at $|\Bet/\Bep|\ge1$ and for very thick flux ropes ($a=0.75$) already at relatively moderate guide field of $|\Bet/\Bep|\ge0.6$. 

Because the guide field decreases only linearly with inverse distance from the line current in the TD model, its effect is expected to be weaker on the Sun. Still, our results suggest that a significant stabilizing effect is provided by a guide field component in solar source regions, because the shear of flare loops indicates that the guide field strength in the flux immediately overlying the erupting flux is often comparable to, or even larger than, the strapping field.

One should be aware that an external toroidal field acts differently if the flux rope is surrounded by vacuum, as in the laboratory studies of the torus instability \cite[e.g.,][]{Myers&al2015, Alt&al2021, Alt&al2023}. In this case, the external field will not change when the unstable flux rope expands, so the initial alignment will be lost. This results in a shielding of the external field via 
a partial rearrangement of the rope current into the surface of the rope, which also changes the forces in the rope. By Lenz's law, the force will act against the further expansion, i.e., stabilizing. However, the details and magnitude of the stabilizing effect appear to 
be different from the case of a plasma environment. The laboratory experiments by \citet{Myers&al2015} and \citet{Alt&al2021, Alt&al2023} are run with an external toroidal field $\Bet$ much larger than $\Bep$ and yet the threshold of the torus instability is estimated to be in the range $n_\mathrm{cr}(h)\sim0.7\mbox{--}1.2$ \citep{Alt&al2021}, near the lower edge of the theoretically predicted range for $\Bet=0$ \citep{Demoulin&Aulanier2010}. Therefore, the stabilizing effect of an external toroidal field must be much weaker if the flux rope is surrounded by a vacuum.

\subsection{Approximation by the Potential Field}\label{ss:results_Bpot}

Figure~\ref{f:n_bet} also shows the decay index at the effective critical height, $\tilde{h}_\mathrm{cr}$, determined with the potential field, $\Bpot$, as an approximation of the external poloidal field, $\Bep$. Two versions of the potential field, computed with the Green function ($B_\mathrm{Green}$) and with the fast Fourier transform ($B_\mathrm{FFT}$), and two components of the former, the full horizontal component ($B_\mathrm{Green,h}$) and the approximate poloidal component ($B_\mathrm{Green,x}, B_\mathrm{FFT,x}$), are considered. 

It is clearly seen that the poloidal component of the potential field, $B_\mathrm{Green,x}$, as well as $B_\mathrm{FFT,x}$, yields a very good approximation of the decay index, with the difference to the true decay index (Equation \eqref{e:n_h_L}) not exceeding $\approx\!0.1$ in our parameter range. 
However, the full horizontal component of the potential field, $B_\mathrm{Green,h}$, which is often used in the literature, yields a good approximation only if the guide field is relatively weak, $|\Bet/\Bep|\lesssim0.6$. In this range, $B_\mathrm{Green,h}$ differs only moderately from $B_\mathrm{Green,x}$.

\subsection{Stabilizing Effect of Line-tying}\label{ss:results_line-tying}

As discussed in Section~\ref{s:review}, line-tying has been suggested to reduce the stability of a coronal flux rope if $\Df>h$, resulting in 
$\ncr<1$, because the major radius of a rising rope must decrease in this range \cite[e.g.,][]{Olmedo&Zhang2010, Alt&al2021}. However, the stabilizing effect of the bending of the rope just above its footprints, which is also required, was not included in the treatment, so the net effect is still uncertain. We expect the effect of the bending near the footprints to increase with increasing $\Df$ in the lower part of the range $\Df>h$. This is based on the increase in the footprint area and on the relative decrease in the middle section of the rope between the footprints (relative to the total rope length) with increasing $\Df$ (and likewise holds for increasing $a$). However, the trend must reverse at some point, 
because line-tying vanishes in the 2D limit $\Df\to\infty$. 

\begin{figure}[t]                                                         
	\centering
        \includegraphics[width=.67\linewidth]{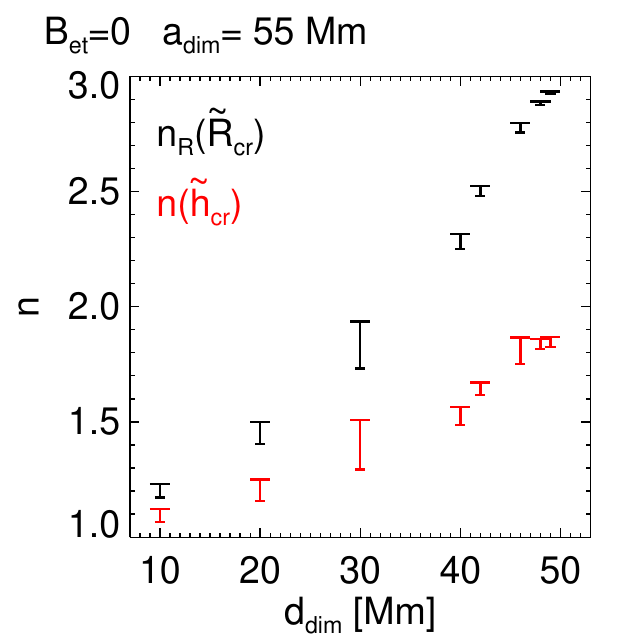} 
	\caption{Critical decay index $n(\tilde{h}_\mathrm{cr})$ (red), transformed to $n_R(\tilde{R}_\mathrm{cr})$ (black), vs.\ dimensional torus center depth $d_\mathrm{dim}$
	for $\Bet=0$, $R_\mathrm{dim}=110$~Mm, and $a_\mathrm{dim}=55$~Mm, showing that $\tilde{n}_{R,\mathrm{cr}}$ approaches the asymptotic value, $n\to3$, 
    when $d_\mathrm{dim}\to\approx 50$~Mm. 
    }
	\label{f:n(R)}
\end{figure}

To see whether the effect of line-tying follows the expected trend, we refer to the free-space case \citep{Kliem&Torok2006}, whose threshold for $\Bet=0$ and $R\gg a$ 
is known (Equation~\ref{e:n_cr3/2-corr}). To this end, we transform the obtained
threshold values $\tilde{n}_\mathrm{cr}$ in terms of height $\tilde{h}$ above the photosphere to the corresponding threshold values in terms of major flux rope radius $R$: $\tilde{n}_{R,\mathrm{cr}}=(\tilde{R}_\mathrm{cr}/\tilde{h}_\mathrm{cr})\,n(\tilde{h}_\mathrm{cr})$, where $\tilde{R}_\mathrm{cr}=\tilde{h}_\mathrm{cr}+d$. In order to make the results for different $\Df$ comparable to each other, we employ dimensional values here, i.e., we fix $R_\mathrm{dim}=110$~Mm (as in TD99) and $a_\mathrm{dim}=55$~Mm (our smallest $a$ at $\Df=1.1$), ensuring a fixed aspect ratio, and vary only the torus depth $d$. For these values, Equation~(\ref{e:n_cr3/2-corr}) yields $n_{R,\mathrm{cr}}=1.26$ independent of $\Df$. That is, for each $\Df$, we compare the threshold of the line-tied flux rope with the threshold of a fully toroidal flux rope in free space, which has the same geometry, so it is governed by the same forces as the line-tied rope at $t=0$. This comparison thus separates the effect of line-tying from the effect of increasingly remote subphotospheric return current in the TD equilibrium. Here we neglect the change in the geometry and forces during the initial relaxation to the numerical equilibrium. This change remains moderate at $\Df=1.1$ and $a_\mathrm{dim}=55$~Mm (Figure~\ref{f:jy}) and is partly compensated at the larger $\Df$ by using the effective height $\tilde{h}$. 

The results, plotted 
in Figure~\ref{f:n(R)}, show (1) an approximate agreement of the numerical threshold $\tilde{n}_{R,\mathrm{cr}}$ with the analytical one for a nearly semicircular geometry ($d_\mathrm{dim}=10$~Mm; $\tilde{h}_\mathrm{cr}/\Df\approx1$), where line-tying should be weakest, and (2) the expected increase in $\tilde{n}_{R,\mathrm{cr}}$ with increasing $d_\mathrm{dim}$ and $\Df$. $\tilde{n}_{R,\mathrm{cr}}$ increases rapidly with $\Df$ and already approaches 3 as $d_\mathrm{dim}\to\approx 50$~Mm ($\Df\approx1.63$; $\tilde{h}_\mathrm{cr}/\Df\approx0.68$). This 
is the asymptotic 
decay index value of the bipole field and, generally, of 
the strapping field of any localized source, i.e., the highest value available, except at poles due to magnetic nulls. Therefore, the numerical result for $\tilde{n}_{R,\mathrm{cr}}$ indicates that the line-tying, which here includes the bending of the top part of the current channel (decreasing $R$) and at its footprints, has a strong, \emph{stabilizing} effect.

On the Sun, this stabilizing effect is counteracted by an increasing force on the coronal current channel from the photospheric boundary (or image current) as $\Df$ increases, which is not correctly modeled by the TD equilibrium. Equilibria that use the image current \citep{Isenberg&Forbes2007, Titov&al2018} must be employed to obtain definite numerical values for the dependence of the threshold on line-tying.

\subsection{Effect of Topology: O-type vs.\ O-X-type Magnetic Structure}\label{ss:results_topology}

For $\Bet=0$, most of our flux rope equilibria are of pure O-type (BPS topology); only the configurations in the top right corner of Figure~\ref{f:jy},
i.e., the equilibria with large $D_\mathrm{f}$ and large $a$, possess an O-X-type (HFT) magnetic structure. These are the most stable ones in our parametric range (Figure~\ref{f:n(D,a)}(b)), and hence have
the smallest values of the sunspot distance $L$ at the marginal stability point (Figure~\ref{f:BP1}). A smaller $L$ not only raises the decay index at the position of the flux rope but also strengthens the bipole field under the current channel relative to the field by the current channel, hence raising the likelihood that an X-type structure forms (see Figures~\ref{f:BP1} and \ref{f:BP2}). Therefore, the formation of O-X-type configurations for large $D_\mathrm{f}$, $a$, and $|\Bet/\Bep|$ is a consequence of higher thresholds in this part of our parameter space (Figures~\ref{f:n(D,a)} and \ref{f:n_bet}), not vice versa. The results in these figures do not permit the conclusion that O-X-type configurations have higher thresholds than pure O-type configurations. 

A consideration of the two configurations adjacent to the topological transition for varying $a$ at fixed $D_\mathrm{f}$ yields some insight into the relevance of the topology for the threshold. If the threshold would be significantly lower for the HFT topology, then the increase in the threshold with increasing $a$ would be counteracted at the transition by a decrease due to the changing topology. However, the data in Figure~\ref{f:n(D,a)} tend to indicate the opposite: in both relevant ranges suggested by Figure~\ref{f:jy} for the numerical equilibria ($a=0.65\mbox{--}0.75$ for $\Df=1.5$ and $a=0.65\mbox{--}0.70$ for $\Df=1.6$), the threshold rises in a particularly strong manner with $a$ at the topological transition, suggesting that the threshold may indeed be \emph{somewhat higher} for O-X-type configurations compared to pure O-type configurations. This apparent contradiction to the theoretical expectation and results \citep{Kliem&Torok2006} is resolved by noticing that all 
numerical equilibria of clear O-X-type in Figure~\ref{f:jy} have developed significant currents also under the HFT. These currents have a net component parallel to the main current channel in the flux rope and hence attract the main current channel. Their poloidal field component acts as an external poloidal field on the main current channel in addition to the one from the subphotospheric bipole; the corresponding Lorentz force is downward. This indeed raises the instability threshold for the numerically relaxed O-X-type configurations in our study. However, it does not contradict the opposite theoretical expectation and result \citep{Kliem&Torok2006}, because the latter refers to configurations without such additional, external current flow. 

An increase in the instability threshold has similarly been found in numerical simulations of CME initiation that include a vertical current sheet under a flux rope \citep{JiangC&al2021}. The flux rope formed at the top of the current sheet was found to be torus stable initially, although the decay index based on the potential field at the formation height was $\approx\!1.7$. However, this potential field was computed under the assumption that there is no current external to the formed flux rope in the coronal volume. 

Overall, our study suggests that the flux rope topology influences the threshold of the torus instability, but it does not allow us to verify the expectation that O-X-type configurations possess a lower threshold than pure O-type configurations. This may be possible when another equilibrium \cite[e.g.,][]{Titov&al2018} is employed. We point out that the twist may also play a role. It is easier for weakly twisted flux ropes to split and erupt in part \cite[as in][]{Gibson&Fan2006}, compared to highly twisted ones. Thus, it is possible that the expected stabilizing effect of line-tying in the bald patches, if any, is weaker for weakly twisted flux ropes, so their critical decay index is less dependent on topology than suggested in \citet{Kliem&Torok2006}.

\section{Conclusions and Discussion}\label{s:concl}

This paper presents a parametric study of the threshold of torus instability (critical decay index) using the model of a force-free, line-tied coronal flux rope equilibrium by \citet{Titov&Demoulin1999} which carries a net current. The ranges $D_\mathrm{f}=1.1\mbox{--}1.63$ for the footpoint distance, $a=0.55\mbox{--}0.8$ for the minor radius of the current channel in the rope (both normalized to the rope's apex height, $h_\mathrm{a}$), and $|\Bet/\Bep|=0\mbox{--}1.73$ for the ratio of external toroidal (guide/shear) and poloidal (strapping) fields at the rope axis are covered. Flatter TD flux ropes ($D_\mathrm{f}>1.63$) were found to be stable due to line-tying and increasingly remote subphotospheric return current. Thinner flux ropes were excluded from the study to avoid the simultaneous occurrence of the helical kink instability. The TD flux rope does not allow studying the 2D--3D geometrical transition independently of other effects, because the subphotospheric return current increasingly deviates from the image of the coronal current channel for a fixed apex height, $h_\mathrm{a}=1$, and an increasing $\Df$.

The approximate analytical TD equilibrium expands in major radius $R$ in the course of relaxation to a numerical equilibrium in our whole parameter range (which is possible even in the unstable cases because the instability has a very small growth rate near the marginal stability point). The expansion depends strongly on the aspect ratio, remaining moderate for our thinnest ropes ($a=0.55$). This indicates that the analytical expressions for the TD flux rope, which are based on the large aspect ratio approximation $R/a\gg1$, still yield an approximately force-free equilibrium for $a\approx0.55$, which corresponds to $R/a\approx2\mbox{--}3.3$ in the considered range of parameters. For our thickest ropes ($a=0.8$; $R/a=1.4\mbox{--}2.3$) the apex height nearly doubles. Additionally, the current density redistributes in the cross section of the current channel in the course of the relaxation, with most of the current flowing significantly below the magnetic axis in our thickest ropes, whereas this height difference remains small for our thinnest ropes. 

We summarize the results for the critical decay index 
at the effective critical height, $\tilde{n}_\mathrm{cr}=n(h\!=\!\tilde{h}_\mathrm{cr})$, 
defined by weighting the height in a wedge at the apex cross section of the current channel with the Lorentz force by the strapping field (Equation~\ref{e:tilde_hcr}). We consider these to be more relevant than the critical decay index values at the magnetic axis of the relaxed numerical equilibria, $n_\mathrm{cr}(h)$; they are also far more consistent with the known analytical and other numerical results. In the considered
range of parameter space we obtain the following results: 
\begin{enumerate} 
\item For zero guide field, $\Bet=0$, the critical decay index is found to lie in the range 
$n(\tilde{h}_\mathrm{cr})=1.1\mbox{--}1.3$
for our smallest $a=0.55\mbox{--}0.65$, where the large aspect ratio approximation is still applicable, and our smallest $\Df=1.1$, where the return current is closest to the image current and line-tying is weakest (Figure~\ref{f:n(D,a)}).
This is consistent with the corresponding cases
studied analytically \citep{Kliem&Torok2006, 
Demoulin&Aulanier2010} 
and with many numerical studies 
(e.g., 
\citealt{Torok&Kliem2007}; 
\citealt{Aulanier&al2010}; \citealt{Zuccarello&al2015, Zuccarello&al2016}), but lower than estimates from solar observations of erupting considerably arched flux ropes. 
For $a>0.65$, 
the critical decay index increases, reaching $n(\tilde{h}_\mathrm{cr})=1.45$ for $a=0.8$. This is likely due to the violation of the large aspect ratio approximation and still lies in the theoretically expected range $\ncr\approx1.1\mbox{--}1.5$. 

\item A guide field, $\Bet\ne0$, can raise the threshold strongly. For example, the threshold 
$n(\tilde{h}_\mathrm{cr})$
rises above 1.5 for $|\Bet/\Bep|\gtrsim0.6$ even for our thinnest ropes with $a=0.55$ and for even 
smaller
$|\Bet/\Bep|\gtrsim0.3$ for thick ropes with $a\gtrsim0.7$ (again, for the smallest $\Df=1.2$ that allows studying the dependence on $|\Bet/\Bep|$ in the TD equilibrium reliably). Such conditions are likely often realized on the Sun. Thresholds $n(\tilde{h}_\mathrm{cr})>2$ are found for $|\Bet/\Bep|\gtrsim1$ in the whole range of $a$ considered. We conjecture that the stabilizing effect of the guide field is the main reason for cases that yielded $n_\mathrm{cr}>3/2$ in the literature \cite[e.g.,][]{Fan&Gibson2007, Fan2010, DuanA&al2019, ChengX&al2020}. However, the effect is likely considerably stronger in the TD equilibrium compared to the conditions on the Sun, 
due to the slow decrease in $\Bet(R)\propto R^{-1}$ in the model. 

\item Line-tying is found to have a stabilizing effect on the torus instability of moderately flat flux ropes that are not 
close to the 2D limit ($h\lesssim\Df$). A common origin of opposite findings in the literature \citep{Olmedo&Zhang2010, Alt&al2021, Filippov2021a} is the omission of 
flux rope bending just above its footprints in these analytical and semianalytical treatments. Line-tying is a strong effect that stabilizes the TD flux rope by itself for $\Df>1.63$ if $\Bet=0$ and above even smaller $\Df$ in the presence of a guide field. For solar flux ropes, it is not yet clear whether it contributes to raising the threshold above the canonical value of 1.5.

\item Most of the numerically relaxed marginally stable flux rope equilibria at $D_\mathrm{f}\le1.5$ or $a\le0.65$ are of pure O (BPS)-type for $\Bet=0$. O-X-type (separator or HFT) equilibria form at larger $\Df$, $a$, or $|\Bet/\Bep|$. However, a secondary current channel develops beneath the separator or HFT in all of them during the relaxation to the numerical equilibrium. This current channel raises the threshold. The theoretically expected lower threshold of O-X-type configurations, compared to pure O-type configurations, in the absence of such a secondary current channel \citep{Kliem&Torok2006} cannot be addressed
by the present investigation. 

\item The potential field based on the Green function yields a very good approximation of the instability threshold for the TD flux rope in our whole range of considered parameters, provided that the poloidal component (horizontal component perpendicular to the flux rope axis) is used. The total horizontal component of the potential field yields a good approximation of the threshold only for a weak guide field, $|\Bet/\Bep|\lesssim0.6$.

\item The assumption of self-similar expansion, used in analytical estimates of the critical decay index for $\Bet=0$, is supported by our study for the relevant case of weakly unstable equilibria (see the appendix). 

\end{enumerate} 

A further parametric study, using an equilibrium with an exact image current below the photosphere, is required 
to substantiate
the above critical decay index values, 
$n(\tilde{h}_\mathrm{cr})$, and to infer their parametric dependence 
in a range 
that covers the conditions in solar eruptions more completely.
The construction of flux rope equilibria based on the Biot-Savart law \citep{Titov&al2018, Titov&al2021, Titov&al2022}
permits such a follow-up investigation. This also has the potential to better link the numerical results to the existing analytical ones by addressing the 3D--2D geometrical transition and provides a 
realistic height dependence of the guide field, $\Bet(h)$.

The comparison with the analytical results can additionally be improved by extending the study to larger aspect ratios $R/a$, which requires suppressing the helical kink instability more effectively. This should be possible by employing 
equilibria that 
have a less uniform twist distribution in the current channel. 
The effect of magnetic topology on the torus instability also needs further study. 

We expect that the strong stabilizing effect of the external toroidal (guide) field, the stabilizing effect of the line-tying for flat, thick flux ropes, the good approximation provided by the poloidal component of the potential field, and the initially mostly nearly self-similar expansion of weakly unstable flux ropes are generally valid results and that most of them stay quantitatively similar to our values for small $\Df$ if an equilibrium with an exact image current is employed. This must be checked. 

The initial expansion of the TD configurations to numerical equilibria indicates a strong role of the distribution of current density in the cross section of the current channel on the critical decay index. Specifically, the thick TD configurations ($a\gtrsim0.7$; $R/a\lesssim2$) relax to numerical equilibria with highly nonuniform current density distribution, resulting in a substantial offset between the magnetic flux rope axis and the effective height, $\tilde{h}$, which refers to the main area of current flow. Very different critical decay index values are obtained from these locations. This offers a further potential explanation for high instability thresholds observationally or numerically inferred in the literature and should receive more attention.

\acknowledgments
We thank Tibor Török, Viacheslav S. Titov and the anonymous referee for helpful comments.
J.C. acknowledges the stimulating atmosphere of the School of Earth and Space Science at University of Science and Technology of China, where part of his work on this paper was done as part of his PhD Thesis. He also acknowledges an invitation and support from the Institute of Physics and Astronomy at the University of Potsdam, Germany, where part of this research was carried out. 
J.C. acknowledges support from the Strategic Priority Research Program of the Chinese Academy of Sciences (grant No. XDB0560000), the National Key R$\&$D Program (grant No. 2022YFF0503003), China's Space Origins Exploration Program (GJ11020215, GJ11020408, GJ11020405), and NSFC grant 41761134088 (in collaboration with DFG).
B.K. acknowledges support by the DFG through grants 392171665 and 520867050. 
R.L. acknowledges support by the National Key R\&D Program of China (2022YFF0503002), the Strategic Priority Research Program of the Chinese Academy of Sciences (XDB0560102), and the NSFC (11925302, 42188101, 42274204).
This research was supported by the International Space Science Institute (ISSI) in Bern through the ISSI International Team project `Understanding the Onset of Solar Eruptions' (ISSI Team project \#24-606).

\appendix
\counterwithin{figure}{section}
\counterwithin{table}{section}
\section{Self-similarity of the Flux Rope After Eruption Onset} 
\label{a:self-similarity}

\begin{figure}[t]                                                          
	\centering
    \includegraphics[width=.75\linewidth]{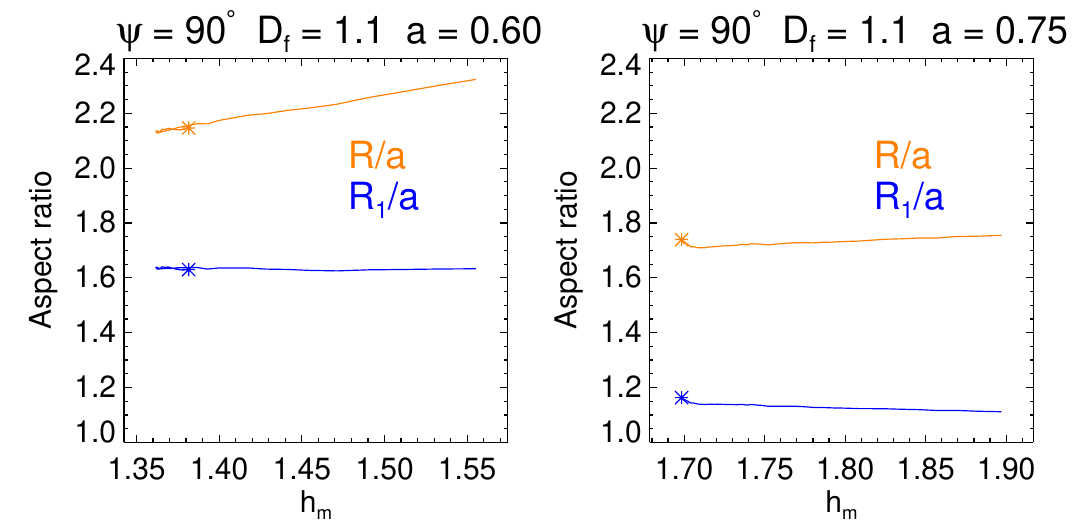}
	\caption{Self-similarity of the initial evolution of the instability, measured by $R/a$ and $R_1/a$, for the marginally unstable configuration shown in Figure~\ref{f:cfl} ($a=0.75$, right panels), and for $a=0.6$ (left panels), both for $\Df=1.1$. 
    The asterisks mark the onset of the instability (shown for $a=0.75$ as a red line in the panels of $L=126$~Mm in Figure~\ref{f:cfl}).
    } 
	\label{f:self-similar}
\end{figure}

The expression for the critical decay index of the toroidal flux rope, Equation~(\ref{e:n_cr3/2-corr}), was obtained using the simplifying assumption that the unstable flux rope evolves self-similarly, $R/a=\mathrm{const}$, justified by the weak (only logarithmic) dependence of the Lorentz self-force
on the aspect ratio \citep{Kliem&Torok2006}. \citet{Alt&al2021} and in part also 
\citet{Olmedo&Zhang2010} adopted this assumption as well. It is known that the main-acceleration phase of CMEs is relatively short, followed by a long propagation phase with typically very little residual acceleration \citep{Zhang&Dere2006, ChengX&al2020} (although cases of substantial additional acceleration after the main acceleration also exist, see, e.g., \citealt{GouT&al2020}). The expansion and contraction between force-free magnetic configurations is self-similar, so that an approximately self-similar expansion of CMEs is expected for the propagation phase, in agreement with observations \cite[e.g.,][]{Kumar&Rust1996}. During the main-acceleration phase, on the other hand, one would not expect a self-similar expansion, unless the forces remain small, which is the case for a very small growth rate of the instability. Therefore, systems initially very close to the marginal stability point may also expand nearly self-similarly during their acceleration phase, especially at the beginning of the acceleration phase when the acceleration has not yet amplified strongly. Here we check this expectation for our marginally unstable configurations during the initial phase of expansion, which is the relevant range for analytical or semi-analytical determinations of the critical decay index. 

Since the relaxation to the numerical equilibrium of the flux rope changes the toroidal shape of the flux rope considerably for some of our configurations, we also consider a modified radius $R_1$ to characterize the expansion in major radius, in addition to $R=h_\mathrm{m}+d$. $R_1$ is the radius of the circle fitted to the apex point and the fixed footpoints of the rope axis, i.e., the radius of the shifted-circle model \citep{ChenJ1989, Olmedo&Zhang2010, Alt&al2021}. The aspect ratio $R_1/a$ changes less than $R/a$ during the initial development of the instability for most of our configurations.

As expected, we find that all our marginally unstable configurations expand approximately self-similarly in the initial phase of instability, which demonstrates the reliability of the assumption in the determination of the critical decay index. For a rise in the flux rope apex by $(0.1\mbox{--}0.4)\,h_\mathrm{m}$ from the onset height of instability, we find the change in $R_1/a$ to stay within $\approx$20\% of the value at instability onset. Most configurations show a change less than 10\%, and some stay within a few percent of the initial value. The latter tend to lie near a diagonal line in Figure~\ref{f:jy}
where the initial shape of the flux rope is closest to a semicircle (apex height after relaxation $h_\mathrm{m}\sim D_\mathrm{f}$). For geometrical reasons, these configurations are expected to adhere to the self-similar assumption much better than initially flat configurations. 

Figure~\ref{f:self-similar} shows the degree of self-similarity for the marginally unstable configuration in Figure~\ref{f:cfl} and for the marginally unstable configuration with $a=0.6$ and $D_\mathrm{f}=1.1$, which adheres best to the assumption. 

\bibliographystyle{aasjournal} 
\bibliography{references_torus3}

\end{document}